\documentclass[]{aa}  

\DeclareUnicodeCharacter{FEFF}{}
\usepackage[]{natbib}
\usepackage[]{fontenc}
\usepackage{ae,aecompl}
\usepackage[dvipsnames]{xcolor}
\usepackage{psfrag,color}
\usepackage[]{inputenc}
\usepackage{graphicx}	
\usepackage{amsmath}	
\usepackage{amssymb}	
\usepackage{longtable}
\usepackage{subfigure}
\usepackage{adjustbox}
\usepackage{ulem}       

\usepackage[]{hyperref}

\newcommand{\hb}{H$\beta$\xspace}

\newcommand{\hi}{H\,{\sc i }}
\newcommand{\hii}{H\,{\sc ii}}
\newcommand{\hei}{He\,{\sc i }}
\newcommand{\heii}{He\,{\sc ii }}
\newcommand{\nii}{N\,{\sc ii }}
\newcommand{\oi}{O\,{\sc i }}
\newcommand{\oii}{O\,{\sc ii }}
\newcommand{\oiii}{O\,{\sc iii }}
\newcommand{\cii}{C\,{\sc ii }}
\newcommand{\ciii}{C\,{\sc iii}}

\newcommand{\nitroi}{[N\,{\sc i}]}
\newcommand{\nitroirec}{N\,{\sc i}}
\newcommand{\feii}{Fe\,{\sc ii}}

\begin{document}

   \title{Excitation mechanisms of {\cii} optical permitted lines in ionized nebulae}

   \subtitle{}

   \author{E. Reyes-Rodr\'iguez
          \inst{1,2},
          J. E. M\'endez-Delgado\inst{3},
          J. Garc\'{\i}a-Rojas\inst{1,2},
          L. Binette\inst{4},
          A. Nemer\inst{5},
          C. Esteban\inst{1,2},
          K. Kreckel\inst{3}
          }

   \institute{Instituto de Astrof\'isica de Canarias, E-38205 La Laguna, Tenerife, Spain \email{ereyes@iac.es} 
         \and
             Departamento de Astrof\'isica, Universidad de La Laguna, E-38206 La Laguna, Tenerife, Spain 
        \and
             Astronomisches Rechen-Institut, Zentrum f\"ur Astronomie der Universit\"at Heidelberg, Mönchhofstraße 12-14, D-69120 Heidelberg, Germany
        \and  
             Instituto de Astronom\'{\i}a, Universidad Nacional Aut\'onoma de M\'exico, A.P. 70-264, 04510 M\'exico, D.F., M\'exico, M\'exico
        \and
            Centre for Astrophysics and Space Science, New York University, Abu Dhabi, UAE
           }

\authorrunning {Reyes-Rodr\'iguez et al.}
\titlerunning {Excitation mechanisms of \cii optical permitted lines in ionized nebulae}
   \date{\today}

 
  \abstract
   {Carbon is the fourth most abundant element in the universe and its distribution is critical to understanding stellar evolution and nucleosynthesis. In optical studies of ionized nebulae, the only way to determine the C/H abundance is by using faint \cii recombination lines (RLs). However, these lines give systematically higher abundances than their collisionally excited counterparts, observable at ultraviolet (UV) wavelengths. Therefore, a proper understanding of the excitation mechanisms of the faint permitted lines is crucial for addressing this long-standing abundance discrepancy (AD) problem.}
   {In this study, we investigate the excitation mechanisms of \cii lines $\lambda\lambda$3918, 3920, 4267, 5342, 6151, 6462, 7231, 7236, 7237 and 9903.}
   {We use the DEep Spectra of Ionized REgions Database (DESIRED) that contains spectra of \hii~regions, planetary nebulae and other objects to analyze the fluorescence contributions to these lines and the accuracy of the atomic recombination data used to model the C$^{+}$ ion.}
   {We find that  \cii $\lambda\lambda$4267, 5342, 6151, 6462 and 9903 arise exclusively from recombinations with no fluorescent contributions. In addition, the recombination theory for these lines is consistent with the observations. Our findings show that the AD problem for C$^{2+}$ is not due to fluorescence in the widely used \cii lines or errors in their atomic parameters, but to other phenomena like temperature variations or chemical inhomogeneities. On the other hand, \cii $\lambda\lambda$3918, 3920, 6578, 7231, 7236, 7237 have important fluorescent contributions, which are inadvisable for tracing the C$^{2+}$ abundances. We also discuss the effects of possible inconsistencies in the atomic effective recombination coefficients of \cii $\lambda\lambda$6578, 7231, 7236 and 7237.}
   {}


   \maketitle
%

\section{Introduction}

 \label{sec:intro}
 
Deep spectra of bright photoionized nebulae are rich in emission lines that allow us to derive their physical conditions and che\-mi\-cal composition. For this reason, detailed knowledge of the atomic processes that give rise to these lines is required. In the first spectroscopic works, some of the brightest lines were associated to the existence of an exotic element, unknown on earth \citep{Huggins:1864} named nebulium \citep{Huggins:1898}. Some years later, however, \citet{bowen27} showed that the unidentified bright emission lines arise from electric dipole-forbidden transitions from  O$^{+}$, N$^{+}$ and O$^{2+}$ ions.

Forbidden lines are produced by magnetic dipole and/or electric quadrupole interactions between metastable states, typically as a result of collisional excitation \citep{Bowen28}, and are being named collisional excited lines (CELs) in such case. Permitted transitions are also common in the nebular spectra, with the \hi and \hei lines being the brightest examples. Many of these lines arise from recombinations of free electrons and are consequently known as recombination lines (hereafter RLs). Permitted lines from heavy elements can also be found \citep{Bowen:1939}, although they can be up to four orders of magnitude weaker than their forbidden counterparts \citep{garciarojasesteban07}.

Nonetheless, not all the forbidden lines originate from collisional excitation, nor are all the permitted lines recombination lines. For example, \nitroi\ $\lambda\lambda 5198, 5200$ or many [\feii] lines in the blue range of the optical spectrum can be produced by continuum pumping fluorescent excitation \citep{rodriguez:1999, Ferland:2012}. The interpretation of these lines in terms of collisional excitation would lead to wrong conclusions about the physical conditions and che\-mi\-cal composition of nebulae. Similarly, fluorescent starlight and/or resonance excitation can also give rise to permitted lines \citep{Bowen:1935, seaton68}, which could lead to an overestimation of the heavy element abundances (relative to hydrogen) when these lines are interpreted as RLs. 

In fact, since the pioneering work by \citet{Bowen:1939} and \citet{wyse42}, it has been known that the heavy element abundances (relative to hydrogen) derived from optical permitted lines are systematically higher than those obtained from forbidden lines \citep[see e.~g.][and references therein]{liu:2006, garciarojasesteban07}, which defines the well known abundance discrepancy (AD) problem. Since this discrepancy can have important implications for understanding the chemical evolution of the universe \citep{peimbertetal17, garciarojasetal19, Maiolino:2019}, it becomes essential to analyze the excitation mechanisms of permitted lines in order to resolve this problem.

\begin{figure}
\centering    
\includegraphics[width=\hsize ]{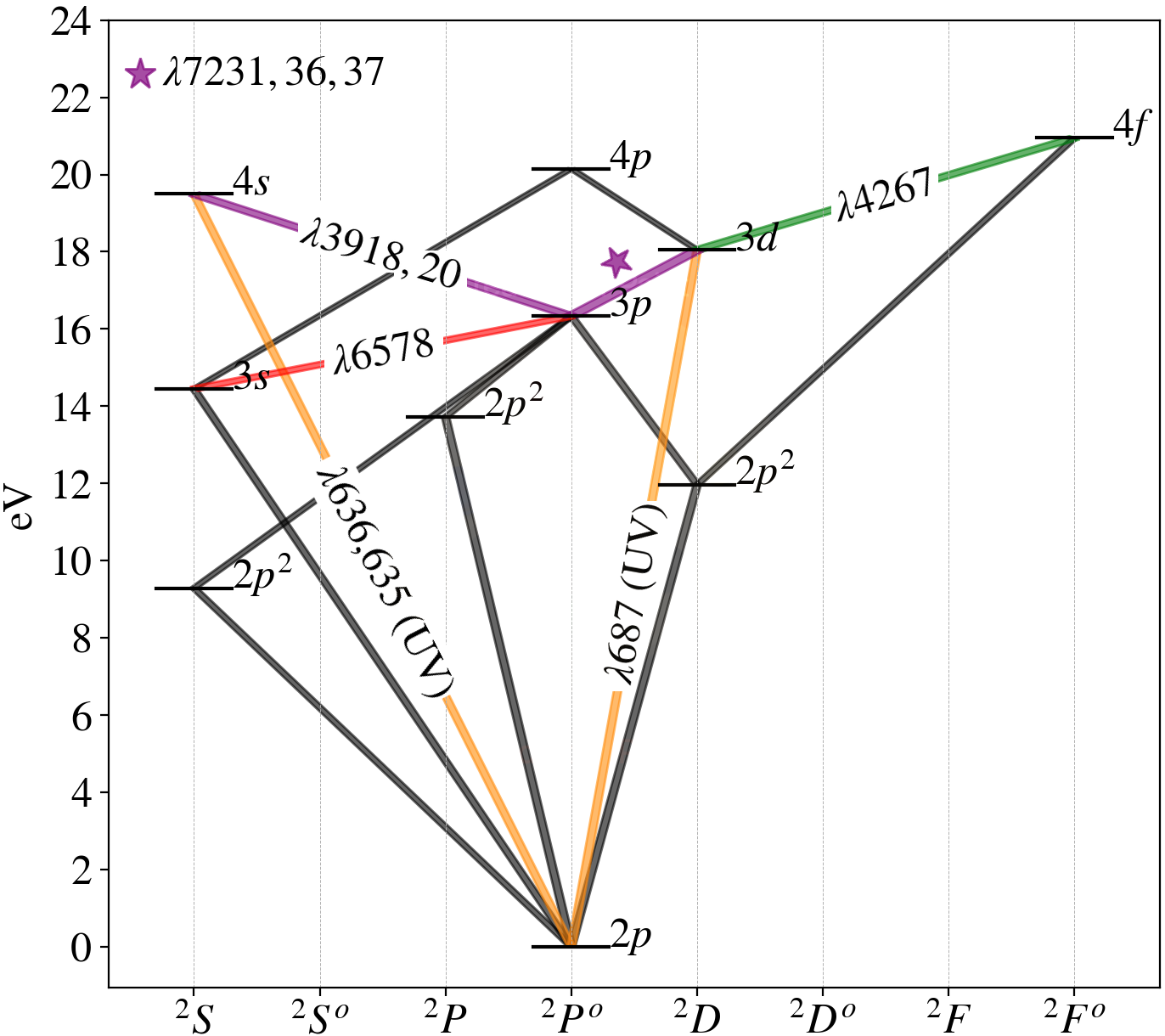}
\caption{Grotrian diagram showing the permitted electronic transitions of the  \cii ion energy levels \citep[adapted from][]{mooremerrill68}. In green we show the transition causing the \cii $\lambda$4267 line where the main excitation mechanism is recombination; purple lines show the transitions behind the \cii $\lambda$$\lambda$3918, 3920 and  $\lambda$$\lambda$7231, 7236, 7237 lines (these last ones marked with a purple star) which are likely to be excited by photon pumping from the continuum (orange transitions). The line we most focus on this project is shown in red. Its corresponding upper level can be seen to be interconnected to the emission of the \cii $\lambda$$\lambda$3918, 3920, $\lambda$$\lambda$7231, 7236, 7237 lines, potentially excited by fluorescence. The values shown on the vertical axis represent the energy level separation from the ground level 2$p$. In grey are shown alternative deexcitation transitions that lead to populating the ground level 2$p$.} 
\label{fig:grotrian_2}
\end{figure}

\citet{escalanteetal12} demonstrated that the ma\-jo\-ri\-ty of observed intensities from the $f$ and $g$ states of \cii, \nitroirec, \nii, \oi, and \oii in the ionized region of the low-excitation planetary nebula IC\,418 could be explained by the available recombination rates currently available. They also found that fluorescence significantly contributed to the excitation of most lines of the $s$, $p$, and $d$ states in \cii, as well as some $p$ and $d$ states of \oii. A similar study that focused on the \nii spectrum of the Orion Nebula was conducted by \citet{escalantemorisset05}. However, there has been no attempt in the literature to generalize results regarding the excitation mechanisms of permitted lines constrained by a large observational sample. This is crucial because the influence of fluorescence on different lines depends on the particular characteristics of the spectral energy distribution (SED) of the ionizing spectrum. Extrapolating conclusions about excitation mechanisms of permitted lines from a study designed for a specific object or a small sample may not be appropriate.

The \cii permitted lines are a key piece for addressing the AD problem since many of them are accessible at optical wavelengths.  In Fig.~\ref{fig:grotrian_2} we show the \cii$-$Grotrian diagram \citep{grotrian:1928} of the optical permitted electronic transitions \citep[adapted from][]{mooremerrill68}. Given the ground state configuration $2p\ ^2P$, high-$L$ energy levels (e.g. $^2F$, $^2G$), will have very low probabilities of being excited by fluorescence \citep{grandi76}, whereas $^2S$, $^2D$ terms can be directly populated through UV continuum photon pumping \citep{grandi76, escalanteetal12}, specially the $4s\ ^2S$, $3d\ ^2D$ states, and potentially excited by $\sim$638 \AA\ and $\sim$687 \AA\ photons, respectively.

Although the probability of direct fluorescent excitation between high-L levels and the ground state can be rather low, cascade interconnections with higher levels could propagate the fluorescence to levels that would otherwise be only populated by recombination. This is the case for the $3p\ ^2P$ level which gives rise to the \cii $\lambda 6578$ line and is interconnected to the $4s\ ^2S$ and $3d\ ^2D$ states via \cii $\lambda \lambda 3918,20$, $\lambda \lambda 7231,36,37$ emission lines. Although unlikely, as radiative $d \rightarrow f$ transitions are rather weak due to their increase in orbital angular momentum, the same principle could apply to other lines considered as RLs, such as \cii $\lambda 4267$. Given the importance of this line, it is necessary to observe and test for the presence of any unforeseen fluorescent effects in a global context.\\

Schematically, a \cii radiative recombination process is represented by: 
\begin{equation}
    \label{eq:recs}
    \text{C}^{2+} + e^- \rightarrow \text{C}^{+} +h \nu,
\end{equation}
where the left side represents the capture of a free electron by the C$^{2+}$ ion, leaving as a result a C$^{+}$ ion and the emission of permitted lines of total energy $h \nu$.
In the case of a fluorescent excitation process:
\begin{equation}
    \label{eq:fluos}
    \text{C}^{+} + h \nu_0 \rightarrow  {^{*}\text{C}^{+}}, 
\end{equation}
where the left side represents the excitation of a C$^{+}$ atom by a photon of energy $h \nu_0$, which gives rise to an excited $^{*}\text{C}^{+}$ ion that will eventually decay to its ground state through a cascade of permitted transitions. This implies that recombination lines will depend on the number of $\text{C}^{2+}$ ions available, whereas those lines produced by fluorescence will depend on the available $\text{C}^{+}$ ions. On the other hand, we should note that the $\text{C}^{2+}$/$\text{C}^{+}$ abundance fraction depends on the ionization parameter of the gas, $U$, and will be correlated with other ionic indicators such as $\text{O}^{2+}$/$\text{O}^{+}$. Therefore, this allows us to observationally establish the relative importance of fluorescent excitation by comparing the line intensity ratio of the {\cii} permitted lines with the degree of ionization of the plasma.

In this work, we analyze the excitation mechanisms of several optical {\cii} permitted lines using the DEep Spectra of Ionized REgions Database \citep[DESIRED,][]{mendezdelgado:2023b}. This allows us to determine the importance of fluorescence excitation on some commonly observed {\cii} lines, by comparing different ionized objects such as {\hii} regions, planetary nebulae (PNe), photoionized Herbig-Haro objects (HHs) and Ring Nebulae (RNe). Furthermore, we test the theoretical predictions of the effective recombination coefficients of \citet{pequignotetal91} and \citet{daveyetal00}, the most widely used to model the {\cii} lines and derive the $\text{C}^{2+}$/$\text{H}^{+}$ abundances. 

This paper is organized as follows: in Sect.~\ref{sec:obs} we describe the observational sample used in this study; in Sect.~\ref{sec:methodology} we give details on the methodology we have followed. The results on the different {\cii} transitions that are the subject of our study are presented in Sect.~\ref{sec:results}. Some discussion on the implications of our findings concerning the abundance discrepancy problem is presented in Sect.\ref{sec:discuss}. Finally our conclusions are summarized in Sect.\ref{sec:conclu}.


\section{Observational sample }
\label{sec:obs}

In order to study the effect of fluorescence on the faint \cii permitted lines, we use the DESIRED observational sample \citep{mendezdelgado:2023b}. This database comprises intermediate-to-high spectral resolution ($R \sim$3,000 to $\sim$33,000) long-slit or echelle spectra compiled from the li\-te\-ra\-tu\-re of about 190 Galactic and extragalactic {\hii} regions as well as Galactic PNe, photoionized Herbig-Haro (HH) objects and ring nebulae around very massive young stars. As the observations were designed to detect very faint emission lines (most of them were obtained with large-aperture – 8–10m – telescopes), this collection of ne\-bu\-lar spectra contains tens or even hundreds of emission lines for each individual object, including multiple faint permitted lines of C, N, O and Ne. A summary of the different spectrographs used to obtain the spectra contained in this database is shown in  Table~\ref{tab:telescopes}. 

\setcounter{table}{0}
\begin{table*}
\caption{ Observatories and spectrographs used to obtain the analysed spectra.}
\label{tab:telescopes}
\centering
\scriptsize
\begin{tabular}{cccccccc}
    \hline
    Spectrograph & Telescope & Diameter [m] & Location \\   \hline 
    OSIRIS & GTC & 10.4 & La Palma, Spain \\ 
    ISIS & WHT & 4.2 & La Palma, Spain \\
    UVES & VLT & 8.2 & Paranal, Chile \\
    FORS2 & VLT & 8.2 & Paranal, Chile \\
    MIKE & Clay Telescope & 6.5 & Las Campanas, Chile \\
    MagE & Baade Telescope & 6.5 & Las Campanas, Chile \\
    HIRES & KECK & 10.0 & Mauna Kea, Hawaii \\
    Blanco echelle & CTIO Blanco & 4.0 & CTIO, Chile\\
    Mayall echelle & KPNO Mayall Telescope & 4.0 & Arizona, USA \\
    \hline
    \end{tabular}
\end{table*}

We hereafter consider all the spectra from the DESIRED sample that present more than two measured optical permitted \cii lines. The list of objects is presented in  Table~\ref{tab:data}. In total, we use 31 spectra of Galactic PNe, 21 of Galactic {\hii} regions, 13 of extragalactic {\hii} regions and 4 of HH objects. More details on the observations referred to in this paper can be found in \citet{mendezdelgado:2023b}, where information on the spectral resolution, wavelength coverage, and the instrument with which each spectra was taken is provided. The lines con\-si\-de\-red in this work and their configuration $-$taken from the Atomic Line List v3.00b4 \citep{vanHoof:2018}$-$ are shown in Table~\ref{tab:transic} and plotted in Fig.~\ref{fig:grotrian_2}.

\setcounter{table}{1}
\begin{table*}
\caption{Electronic configurations and energy levels of the different \cii transitions analysed in this work.}
\label{tab:transic}
\centering
\scriptsize
\begin{tabular}{cccccccc}
    \hline
    Lab. wavelength [Å] & Configuration & Term & Level energy [eV] \\  
    \hline 
    3918.978 & 2s$^2$.3p--2s$^2$.4s & $^2$P$^{\mathrm{o}}$--$^2$S & 16.331742 -- 19.494538  \\
    3920.693 & 2s$^2$.3p--2s$^2$.4s & $^2$P$^{\mathrm{o}}$--$^2$S & 16.333124 -- 19.494538  \\
    4267.001 & 2s$^2$.3d--2s$^2$.4f & $^2$D --$^2$F$^{\mathrm{o}}$ & 18.045808 -- 20.950642 \\ 
    4267.183 & 2s$^2$.3d--2s$^2$.4f & $^2$D --$^2$F$^{\mathrm{o}}$ & 18.045985 -- 20.950695 \\
    4267.261 & 2s$^2$.3d--2s$^2$.4f & $^2$D --$^2$F$^{\mathrm{o}}$ & 18.045985 -- 20.950642 \\
    5342.500  & 2s$^2$.4f--2s$^2$.7g & $^2$F$^{\mathrm{o}}$--$^2$G & 20.950695 -- 23.270770 \\
    6151.530  & 2s$^2$.4d--2s$^2$.6f & $^2$D --$^2$F$^{\mathrm{o}}$ & 20.844773 -- 22.859716 \\
    6461.950  & 2s$^2$.4f--2s$^2$.6g & $^2$F$^{\mathrm{o}}$--$^2$G & 20.950695 -- 22.868793 \\
    6578.048  & 2s$^2$.3s--2s$^2$.3p & $^2$S --$^2$P$^{\mathrm{o}}$ & 14.448827 -- 16.333124  \\
    7231.340  & 2s$^2$.3p--2s$^2$.3d & $^2$P$^{\mathrm{o}}$--$^2$D  & 16.331742 -- 18.045807 \\
    7236.420  & 2s$^2$.3p--2s$^2$.3d & $^2$P$^{\mathrm{o}}$--$^2$D  & 16.331742 -- 18.045807 \\
    7237.170  & 2s$^2$.3p--2s$^2$.3d & $^2$P$^{\mathrm{o}}$--$^2$D  & 16.331742 -- 18.045807 \\
    9903.890  & 2s$^2$.4f--2s$^2$.5g & $^2$F$^{\mathrm{o}}$--$^2$G & 20.950695 -- 22.202226 \\
    \hline
    \end{tabular}
\end{table*}

\begin{figure}
\centering
\includegraphics[width=\hsize]{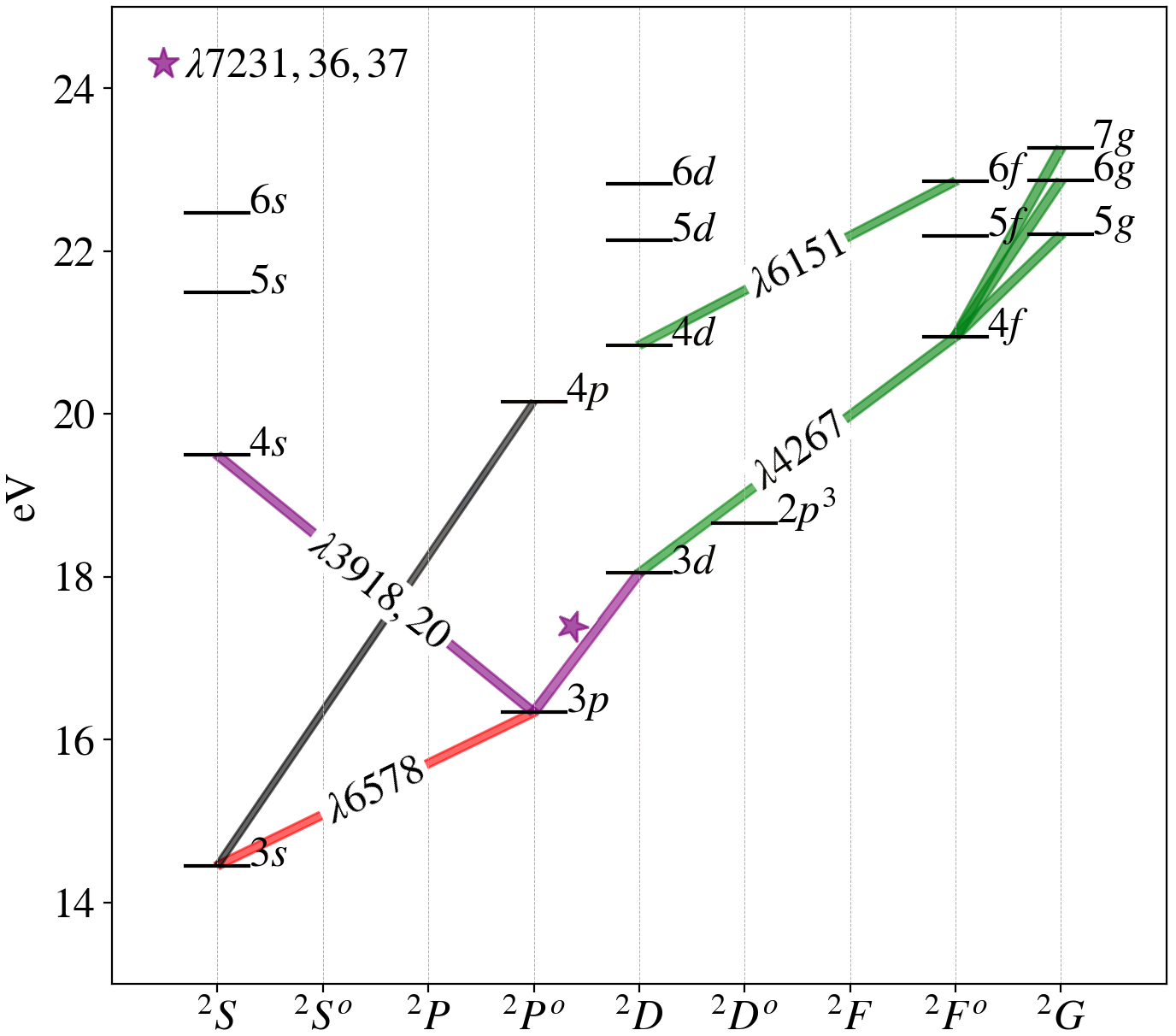}
\caption{Grotrian diagram showing permitted electronic transitions between the energy levels of the \cii ion \citep[adopted from][]{grandi76}. In green we show the transitions which generate the \cii $\lambda$4267, $\lambda$5342, $\lambda$6151, $\lambda$6462 and $\lambda$9903 recombination lines; in purple we show the transitions causing the \cii $\lambda\lambda$3918, 3920, $\lambda\lambda$7231, 7236, 7237 lines that are likely excited by photon pumping from the continuum. The line that focus our attention is shown in red. Its upper source level is interconnected with the emission of the \cii $\lambda\lambda$3918, 3920, $\lambda\lambda$7231, 7236, 7237 lines. The values in $eV$ on the vertical axis represent the energy level separations from the ground level. Notice that the 4f$-$7g, 4f$-$6g and 4f$-$5g transitions shown in green give rise to the \cii $\lambda$5342, $\lambda$6462 and $\lambda$9903 lines, respectively, and subsequently to the emission of the $\lambda$4267 line.}
\label{fig:grotrian}
\end{figure}


\section{Methodology}
\label{sec:methodology}

In order to directly analyze the excitation mechanisms of the various transitions shown in Table~\ref{tab:transic}, we proceed to compare the different line intensity ratios with the degree of ionization as traced by the parameter $P$ \citep{pilyugin01}, which is defined as follows:

\begin{equation}
\label{eq:P}
P = \frac{I([\text{O~III}] \lambda \lambda 4959+5007)}{I([\text{O~III}] \lambda \lambda 4959+5007)+I([\text{O~II}]  \lambda \lambda 3726+3729)}.
\end{equation}

Interestingly, our sample extends over the full range of ionization degree, from $P\simeq 0$ up to $P\simeq 1$. We are aware that this parameter is slightly sensitive to both electron density and temperature, however, we have utilized the model sample from BOND \citep{valeasarietal16} to confirm the good correlation between $P$ and the degree of ioni\-zation (given by O$^{2+}$/O). Additionally, it is a more easi\-ly observable ratio than e.~g. \heii/\hei, which is not observable in most \hii\ regions, or \hei/\hi which can be affected by uncorrected \hei stellar absorptions. Given the high spectral resolution of some observations reported by DESIRED, some multiplets could actually be resolved into their different component lines. This is the case for instance for the $3p\ ^2P^o-4s\ ^2S$, $3d\ ^2D-4f\ ^2F^o$, $3p\ ^2P^o-3d\ ^2D$ multiplets. In order to establish an auto-consistent com\-pa\-ri\-son throughout the whole data sample, we have summed up the intensities of the different components to simulate a blend of the multiplet lines. 

As mentioned in Sect.~\ref{sec:intro}, the intensity of a line is proportional to the density of emitting ions. If we compare two recombination lines, the intensity of each line will be proportional to the C$^{2+}$ density multiplied by the electron density ($n_{\rm e}$) and the electron temperature ($T_{\rm e}$) at a power close to $-1$. Therefore, the line intensity ratio can be expected to be fixed by the atomic recombination probability ratios.\\ 

On the other hand, if one line is excited by recombination while the other arises mostly from fluorescence, the line intensity ratio will depend on the  $\text{C}^{2+}$/$\text{C}^{+}$ abundance ratio. Therefore, the line intensity ratio is expected to correlate with $P$, although the correlation can become more complex as some hard-to-know factors such as the effective temperature of the ionizing source or the optical depth may also play a significant role. The most complex case will arise when comparisons are made with two lines arising from fluorescent excitation. This analysis could for instance reveal the prevalence of one fluorescence channel with respect to another as a result of differences in their physical conditions or to their more direct interconnection.

In order to compare the observed line intensity ratios with theoretical predictions, we test the effective recombination coefficients of  \citet{pequignotetal91} and \citet{daveyetal00}. Briefly, \citet{pequignotetal91} calculated the total and effective radiative \cii recombination coefficients by considering an electron temperature that ranges from $\sim$ 5,000 K  to $\sim$ 20,000 K for a low-density (optically thin) plasma. On the other hand, \citet{daveyetal00} considered the effective recombination coefficients for \cii transitions for a temperature that ranges from 5,000 K to 20,000 K and at a constant electron density of 10$^4$ cm$^{-3}$. They obtained bound-bound and bound-free radiative estimates that incorporate both radiative and dielectronic recombination effects. In addition, they also include the effects of electronic collisions which can induce excited states.


\section{Results from our analysis}
\label{sec:results}

In this Section we show the different \cii line intensity ratios as function of $P$. In all figures we use different markers in order to distinguish line intensity ratios from Galactic and extragalactic {\hii} regions, PNe, RNe and HHs. We also consider two color bands with the theoretical predictions of \citet{pequignotetal91} and \citet{daveyetal00} covering the range from 4,000$-$15,000 K at a density of 10$^3$\,cm$^{-3}$. The code {\sc PyNeb} v1.1.16 Python package \citep{luridianaetal15} was used to calculate these line intensity ratios. These color bands allow us to distinguish whether the observed dispersion may be due to temperature variations or to fluorescence effects. In all cases we adopt the theoretical ratios predicted under case\,B for the \cii levels. Note, however, that those lines arising from levels of $^2F^o$, $^2G$ terms are case-independent.

We begin by analyzing the $^{2}F^{o} {-} ^{2}G$ and $^{2}D {-} ^{2}F^{o}$ transitions since in principle they are the least likely transitions to be excited through continuum pumping. Their intensities will be subsequently compared to the \cii\ $\lambda 4267$ line, since its observed intensity is by far the most frequently used to estimate the abundance ratio of C$^{2+}$/H$^{+}$ \citep{Peimbert:1995, estebanetal05, estebanetal14, Skillman:2020}. Afterwards, we will analyze the $^{2}P^{o} - ^{2}D$ and $^{2}P^{o} - ^{2}S$ transitions and their relationship to the \cii $\lambda 4267$ line. Continuum pumping via fluo\-rescent excitation is likely to contribute greatly to the $^{2}S$ and $^{2}D$ levels. Finally, we will focus on the excitation mechanism of the \cii $\lambda 6578$ line that arises from the $^{2}S-^{2}P^{o}$ transition and which can be contaminated through secondary fluorescent channels.

\subsection{\cii $^{2}F^{o} {-} ^{2}G$ and $^{2}D {-} ^{2}F^{o}$ transitions: $\lambda\lambda$5342, 6462, 9903 and $\lambda\lambda$4267, 6151}
\label{subsec:recs_lines}

Direct excitation from the ground state to the $^{2}G$ or $^{2}F^{o}$ levels are strongly forbidden given that $\Delta L=3$ and $\Delta L=2$, res\-pectively. Therefore, in order to get important continuum pumping contribution to the intensity of lines such as \cii $\lambda\lambda 4267, 6151, 5342, 6462, 9903$ (see Fig.~\ref{fig:grotrian}), an interconnection via a fluorescent channel is required. Such channel could be due to a metastable intermediate level that promotes self-absorption or to cascade-decays from upper levels via continuum pumping excitation. In the case of transitions arising from the $^{2}G$ levels, the last scenario would require decaying through two intermediate levels in order to preserve $\Delta L=1$. As the aforementioned lines arise from very high excited levels with energies close to the ionization threshold, this process appears rather unlikely.\\

\begin{figure}
\centering
\includegraphics[width=\hsize]{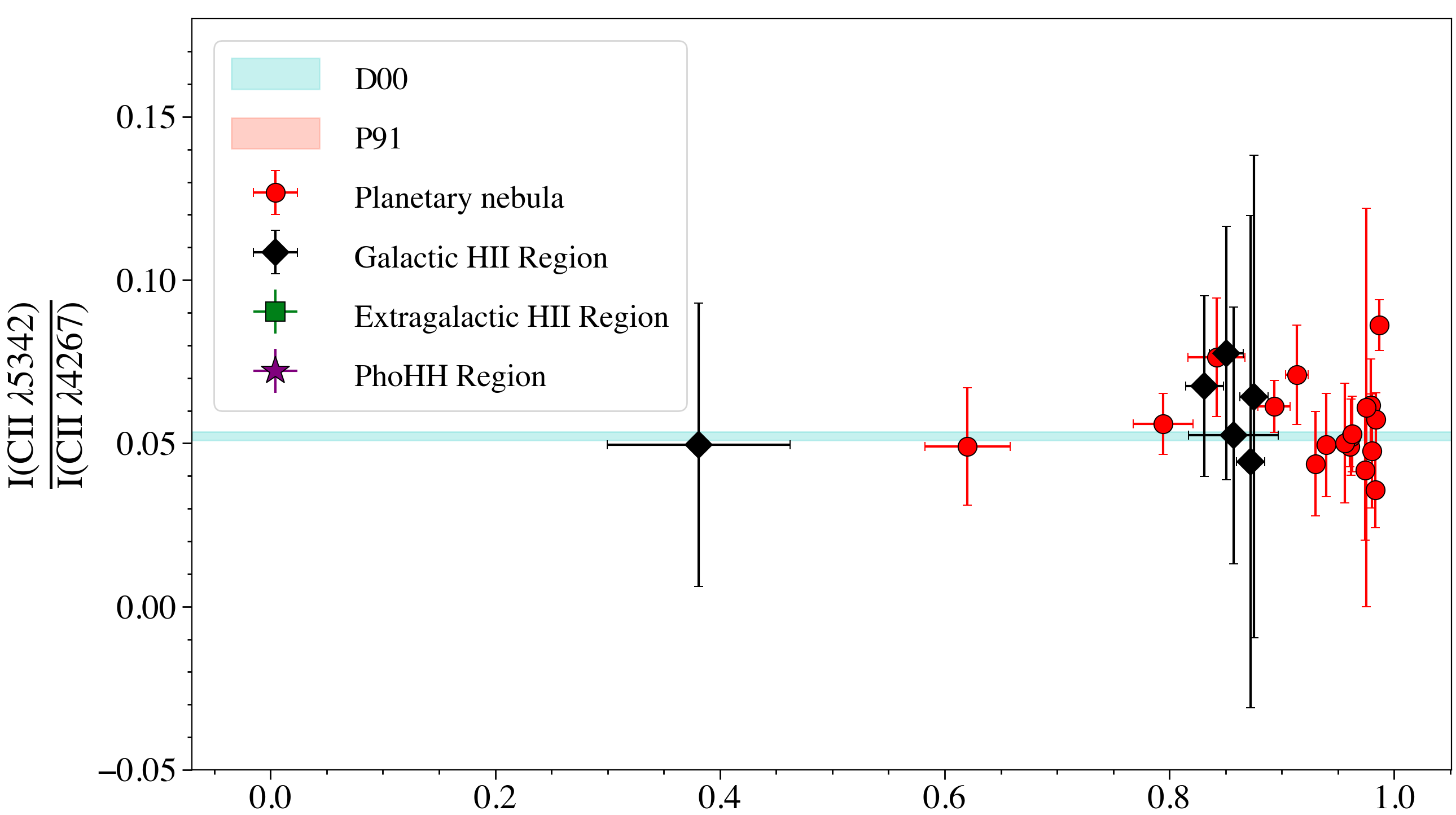}
\includegraphics[width=\hsize]{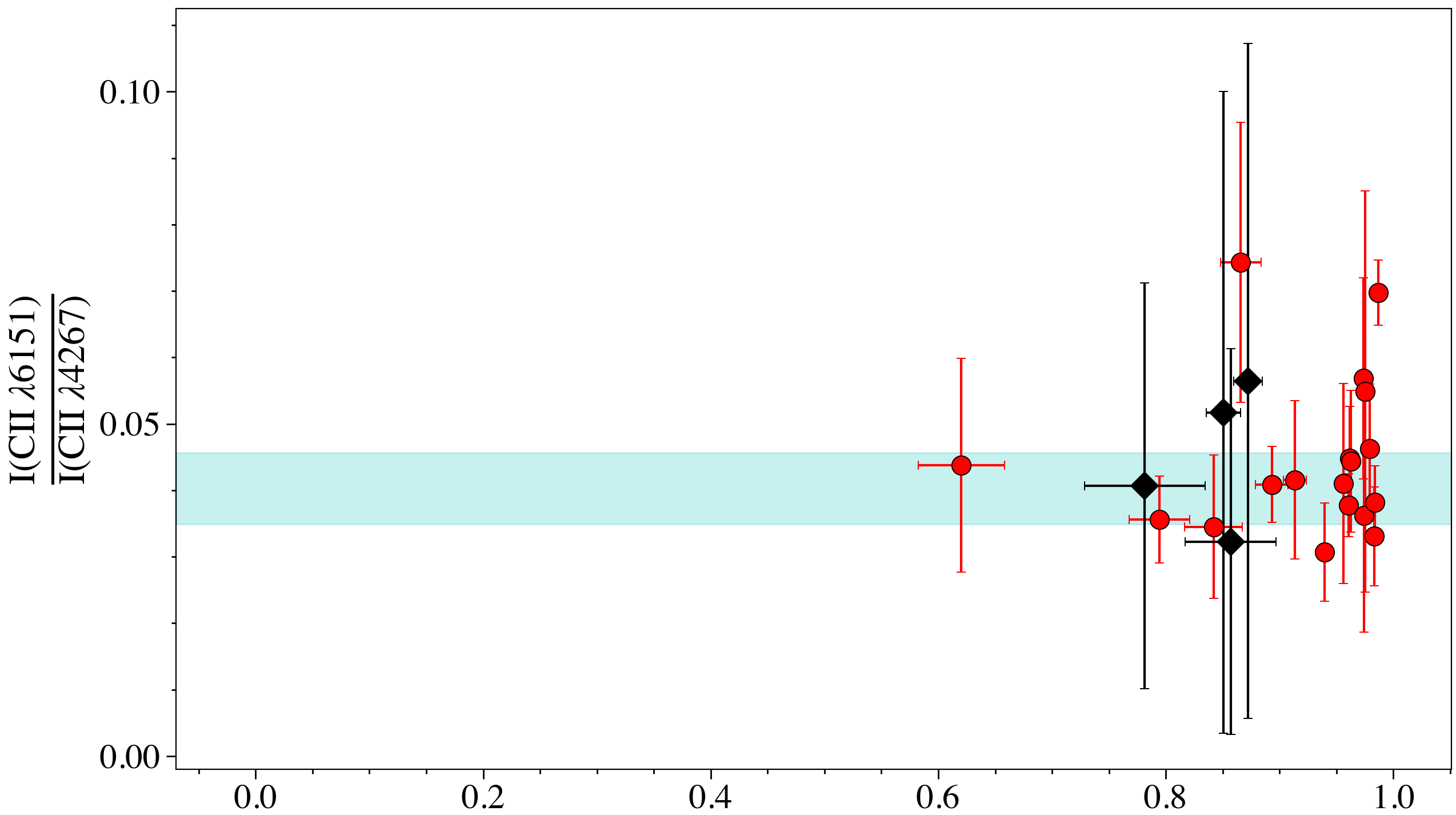}
\includegraphics[width=\hsize]{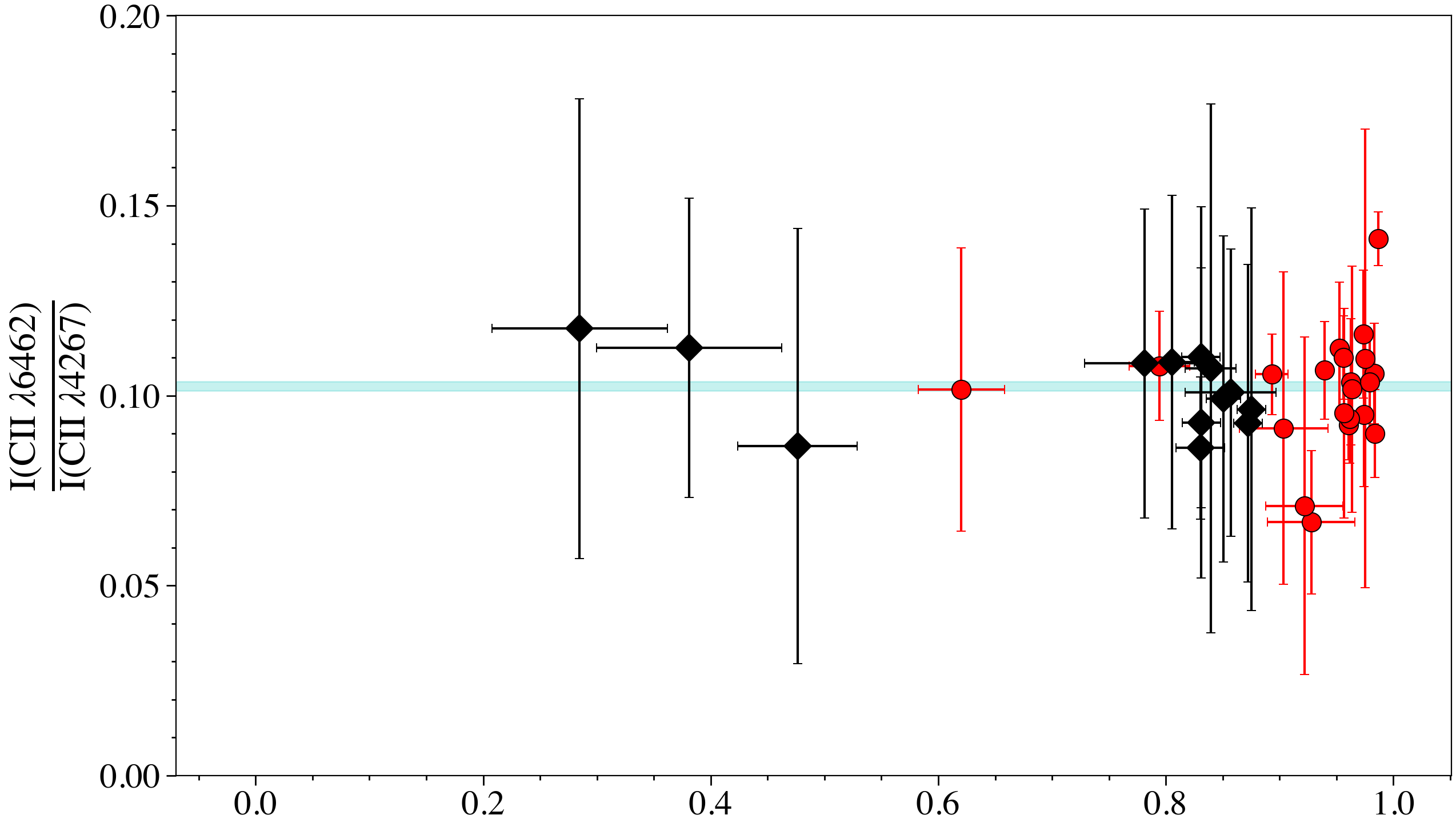}
\includegraphics[width=\hsize]{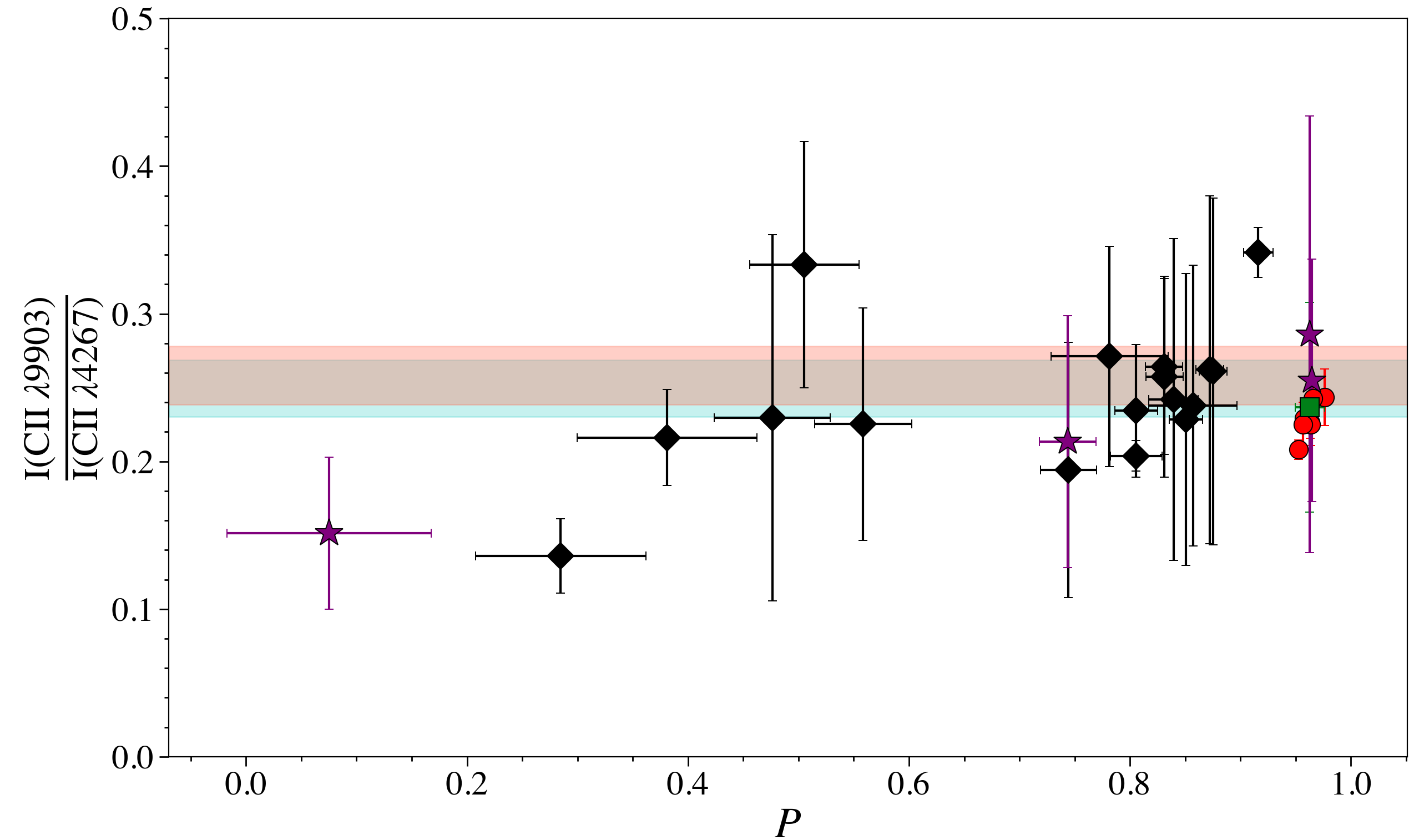}
\caption{From top to bottom, I(\cii $\lambda$5342)/I(\cii $\lambda$4267), I(\cii $\lambda$6151)/I(\cii $\lambda$4267), I\ (\cii $\lambda$6462)/ I(\cii $\lambda$4267) and I(\cii $\lambda$9903)/I(\cii $\lambda$4267) ratios of all analyzed regions: Galactic PNe (red circles), Galactic {\hii} regions (black diamonds), extragalactic {\hii} regions (green squares) and photoionized Herbig-Haro objects (purple stars) as a function of the degree of ionization $P$. The red band shows the theoretical ratio expected for nebular conditions consisting of a temperature in the range 4,000$-$15,000 K and a density 10$^3$ cm$^{-3}$ assuming the atomic data of \citet{pequignotetal91} (P91, only available for {\cii} 4267 and 9903) while the blue band shows the theoretical ratio expected when adopting the atomic data of \citet{daveyetal00} (D00).}
\label{fig:tipos}
\end{figure}

\setcounter{table}{2}
\begin{table*}
\caption{Observed line intensity ratios from the \cii $^{2}F^{o} {-} ^{2}G$ and $^{2}D {-} ^{2}F^{o}$ transitions which are compared to theoretical predictions where the temperature covers the range 4,000K to 15,000K.}
\label{tab:RLS_OBS_values}
\centering
\scriptsize
\begin{tabular}{cccccccc}
    \hline
    Line intensity ratio &  Predicted values & Predicted values  & Observed values & $N$\\
    & \citep{pequignotetal91} & \citep{daveyetal00} & & \\
    
    \hline
I(\cii $\lambda$5342)/I(\cii $\lambda$4267)&-&0.051-0.054 &$0.058 \pm 0.013$& $24$\\
I(\cii $\lambda$6151)/I(\cii $\lambda$4267)&-&0.035-0.046&$0.044 \pm 0.012$& $22$\\
I(\cii $\lambda$6462)/I(\cii $\lambda$4267)&-&0.104-0.101&$0.107 \pm 0.018$& $34$\\
I(\cii $\lambda$9903)/I(\cii $\lambda$4267)&0.278-0.238 & 0.269-0.230& $0.224 \pm 0.031$& $29$\\
    \hline
    \end{tabular}
\end{table*}

In Fig.~\ref{fig:tipos} we show the line intensity ratios I(\cii $\lambda$5342)/I(\cii $\lambda$4267, I(\cii $\lambda$6151)/I(\cii $\lambda$4267), I(\cii $\lambda$6462)/I(\cii $\lambda$4267) and I(\cii $\lambda$9903)/I(\cii $\lambda$4267) as a function of $P$. The panels are ordered according to the energy level that gives rise to the line which appears as numerator. The top panel always corresponds to the highest energy level (e.g., \cii $\lambda5342$: 23.27\,eV). It should be noted that among the lines considered, \cii $\lambda4267$ is the line which arises from the lowest energy level in Fig.\,\ref{fig:grotrian_2} (20.95\,eV). In all cases, Fig.~\ref{fig:tipos} shows rather constant line intensity ratios with respect to $P$ values. The observed values and their dispersion are shown in Table~\ref{tab:RLS_OBS_values}. 

A quick glance at Fig.~\ref{fig:tipos} reveals that the data follows the expected trend for lines produced by recombination. Furthermore, Table~\ref{tab:RLS_OBS_values} reveals good consistency with the theoretical predictions for these transitions from both  \citet{pequignotetal91} and \citet{daveyetal00}. The two outliers shown in the bottom panel of Fig.~\ref{fig:tipos} are HH~204 \citep{mendez-delgado21b} and M~16 \citep{garciarojasetal06}. In the first case, it is explained by the partial blend that occurs between the \cii $\lambda4267$ emission from HH~204 and the ne\-bu\-lar emission from M~42. In the case of M~16, \citet{garciarojasetal06} report the presence of telluric absorption bands that affect the \cii $\lambda9903$ line, which explains the observed behaviour.

\subsection{\cii $^{2}P^{o} {-} ^{2}D$ and $^{2}P^{o} {-} ^{2}S$ transitions: $\lambda$7231+36+37 and $\lambda$3918+20}\label{subsec:3918_7231}

In addition to recombination processes, the \cii $^{2}D$ and $^{2}S$ levels can be populated through continuum pumping fluorescence channels which connect to the ground level, as indicated in Fig.~\ref{fig:grotrian}. These levels give rise to the following lines $\lambda$3918+20 ($3p\ ^{2}P^{o} - 4s\ ^{2}S$) and $\lambda$7231+36+37 ($3p\ ^{2}P^{o} - 3d\ ^{2}D$), commonly observed in deep optical spectra. By comparing the line intensity ratio of these lines with $\lambda$4267 (or any other recombination line, see Sect.~\ref{subsec:recs_lines}) it is possible to determine the fraction due to fluorescence. 

\begin{figure}
\centering
    \includegraphics[width=\hsize]{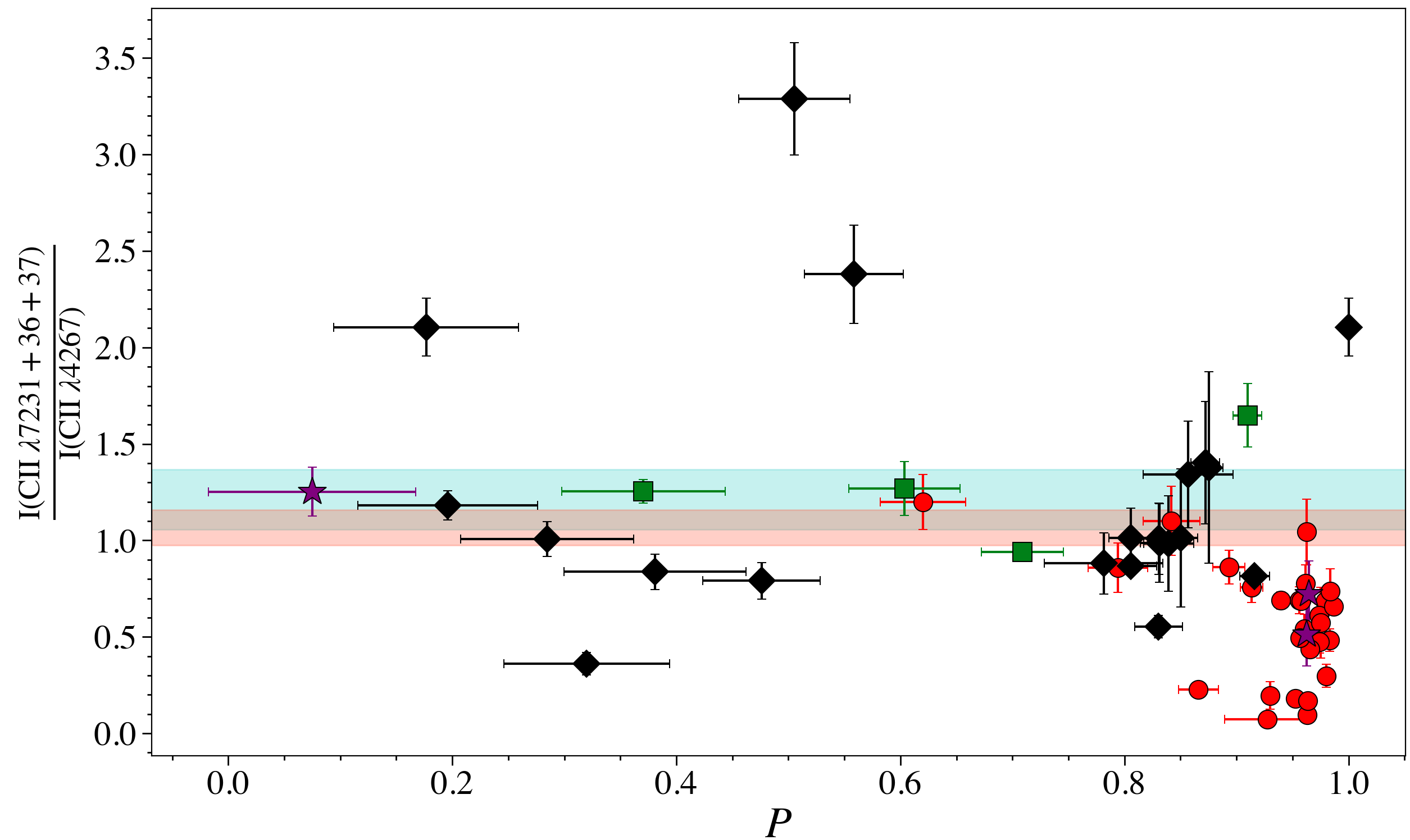}
    \includegraphics[width=\hsize]{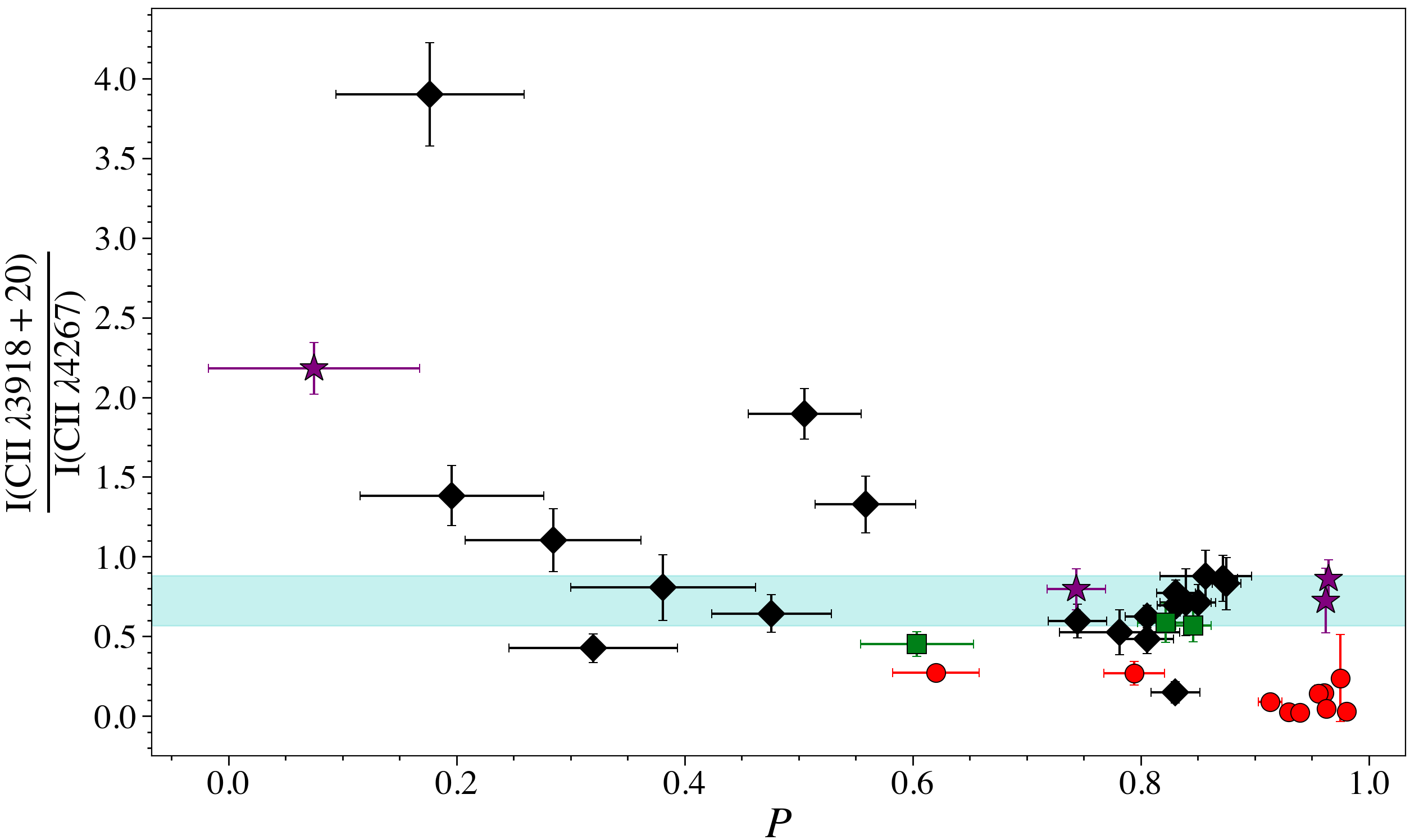}
\caption{I(\cii $\lambda$7231+36+37)/I(\cii $\lambda$4267) ratio (top panel) and I(\cii $\lambda$3918+20)/I(\cii $\lambda$4267) ratio (bottom panel) of all analyzed regions with respect to the degree of ionization $P$. Symbols and color bands are the same as in Fig.~\ref{fig:tipos}. Notice that \citet{pequignotetal91} do not include computations for the \cii $\lambda\lambda$3918, 3920 lines.}
\label{fig:all8}
\end{figure}

In Fig.~\ref{fig:all8} we show the observed I(\cii $\lambda$7231+36+37)/I(\cii $\lambda$4267) and I(\cii $\lambda$3918+20)/I(\cii $\lambda$4267) line intensity ratios. Unlike what is shown in Fig.~\ref{fig:tipos}, these line intensity ratios show a trend with the ionization degree of the gas, as represented by $P$. This means that a fraction of the emission of \cii $\lambda$3918+20 and $\lambda$7231+36+37 is proportional to the ionic density of C$^{+}$ rather than C$^{2+}$, indicating a general contribution from fluorescence. In Fig.\,\ref{fig:all8}, I(\cii $\lambda$3918+20)/I(\cii $\lambda$4267) line intensity ratio shows a rather clear trend with respect to $P$. This seems to indicate that recombination in \cii $\lambda$7231+36+37 could be more important than for \cii $\lambda$3918+20, although it may be not the dominant excitation mechanism. It seems that for some PNe with the highest degree of ionization, where recombination dominates over fluorescence as the C$^{+}$/C$^{2+}$ is minimum, the line ratios saturate to a minimum value. The wide dispersion shown in these figures is due to the dependence of fluorescence on both the optical depth and the effective temperature of the ionizing source. These parameters on the other hand are expected to display a wide range of values among \hii~regions, PNe and HHs.

It is interesting to note in the top panel of Fig.~\ref{fig:all8} that a signi\-ficant fraction of objects show I(\cii $\lambda$7231+36+37)/I(\cii $\lambda$4267) values below the recombination predictions of the atomic models of both \citet{pequignotetal91} and \citet{daveyetal00}, assuming case\,B and gas temperatures between 4,000$-$15,000 K. This could be due to different factors: (i) the adopted effective recombination coefficients underestimate the recombination emissivity of \cii $\lambda 4267$, (ii) the gas temperature is much lower than 4000~K, (iii) disturbance on the characteristics of the atomic levels (broadening, lifetime, etc) due to fluorescence which would change the coefficients of all those high excited states when solving their level populations equations \citep{Nemer:2019}, (iv) when the population of C$^{+}$/C$^{2+}$ is low enough, the 687\AA~photon pumping from ground level to the $3d\ ^{2}D$ level is negligible, practically reaching ``case A'' optical conditions for this \cii transition or (v) the adopted effective recombination coefficients overestimate the recombination emissivity of \cii $\lambda$7231+36+37.

The results of Sect. \ref{subsec:recs_lines} discard the first scenario since Fig.~\ref{fig:tipos} shows an excellent agreement between the predicted values of \cii $\lambda$4267 with respect to the other lines that are produced by pure recombination. The second scenario is discarded by the fact that even under the extremely low temperature conditions of $\sim 500$ K the prediction is $\lambda$7231+36+37/$\lambda$4267$ \sim 0.9$. 

Since fluorescence is a radiation-related process, a theoretical model that includes radiation transport is needed to explore hypotheses (iii) and (iv). We use the code Cloudy \citep{Ferland:2017} to simulate the Orion Nebula, the \hii~ region known in the greatest detail. The central source is taken from Kurucz stellar atmosphere atlas \citep{Kurucz:1991} with $T_{eff}$ = 39,600 K which depicts the detailed SED of an OVI star as the ionizing source of the nebula with a surface flux of hydrogen-ionizing photons of  $10^{13}$ cm$^{-2}$ s$^{-1}$. The model is defined with constant gas pressure, particle density $10^4$ cm$^{-3}$ and Orion Nebula abundances and grains. We use the atomic data from Chianti database \citep{Landi:2012} which include 158 energy levels for the C$^+$ ion. For more details please refer to Cloudy test suite {\it orion\_hii\_open}. We predict the intensity of the {\cii} $\lambda$3918+20, $\lambda$4267, $\lambda$6578, $\lambda$7231+36+37, and $\lambda$9903 lines, and calculate their ratios obtaining the results shown in Table~\ref{tab:models}. In this table we also compare the predicted ratios with the observed values obtained by \citet{estebanetal04} and the recombination predictions from \citet{daveyetal00} . 

\setcounter{table}{3}
\begin{table}
\centering
\scriptsize
\caption{\small \cii line ratios obtained from a photoionization models simulating Orion's dust and chemical abundances, the observed values from \citet{estebanetal04}. and the predicted values from recombination by \citet{daveyetal00}.}
\label{tab:models}
\begin{tabular}{cccccccc}
    \hline
    Ratio & Model & Observations & Rec.  \\   
     &  &  &  prediction \\   \hline
    $P$ & 0.8 & 0.86 &-&\\
    I(\cii $\lambda$7231+36+37)/I(\cii $\lambda$4267) & 0.40 & 1.34 & 1.19\\
    I(\cii $\lambda$3918+20)/I(\cii $\lambda$4267) & 0.68 &0.88& 0.05 \\
    I(\cii $\lambda$9903)/I(\cii $\lambda$4267) & 0.25 & 0.24 & 0.25 \\
    I(\cii $\lambda$6578)/I(\cii $\lambda$4267) & 0.42 & 1.17 & 0.77 \\
    \hline
    \end{tabular}
\end{table}

From Table~\ref{tab:models} it is clear that by considering radiation transport in the photoionization model, we can get values both above and below the recombination predictions. However, the ratios predicted by the model differ significantly from the observed values, except for the case of the I(\cii $\lambda$9903)/I(\cii $\lambda$4267) line ratio, which involves two lines purely excited by recombination (see Sect.~\ref{subsec:recs_lines}). This highlights the difficulty of properly modeling the line fluxes when fluorescent excitation plays an important role. Creating a detailed model for each of the studied regions here is beyond the scope of this article. Such models would require comprehensive information about the effective temperature of the ionizing sources, which is not available in many cases. It is important to mention that effect (iii) is difficult to differentiate from (iv). In practical terms, continuum pumping from the ground level to the $3d\ ^{2}D$ level is a fluorescent effect inherently related to the optical depth of C$^{+}$.

The previous discussion shows that the observed trends in Fig.~\ref{fig:all8}, which includes the presence of some points below the theoretical predictions of recombination, is an effect of fluorescence. However, we cannot rule out the possibility of having some overestimation of the recombination emissivity due to the effective recombination coefficients of \cii $\lambda$7231+36+37 as mentioned in (v). Overestimation in the effective recombination coefficients related to lines \cii $\lambda$7231+36+37 could cause an underestimation of the derived fluorescent contribution to these same lines. 

\subsection{\cii $^{2}S{-} ^{2}P^{o}$ transition: $\lambda$6578}\label{subsec:6578}

We will now focus on analyzing the excitation mechanism of the $3p^{2}P^{o}$ level which gives rise to the \cii $\lambda$6578 line by decaying to the $3s\ ^{2}S$ level. The $3p\ ^{2}P^{o}$ level does not have a direct continuum pumping channel from the ground level. However, the transitions $3p\ ^{2}P^{o} - 4s\ ^{2}S$ and $3p\ ^{2}P^{o} - 3d\ ^{2}S$, which give rise to the \cii $\lambda$3918+20 and $\lambda$7231+36+37 lines could propagate the fluorescent effects to the $3p\ ^{2}P^{o}$ level. If recombination could dominate over fluorescence in exciting \cii $\lambda$6578 line, at least for nebulae within a certain range of degree of ionization, this line might turn out a good tracer of C$^{2+}$ abundance.

\begin{figure}
\centering
    \includegraphics[width=\hsize]{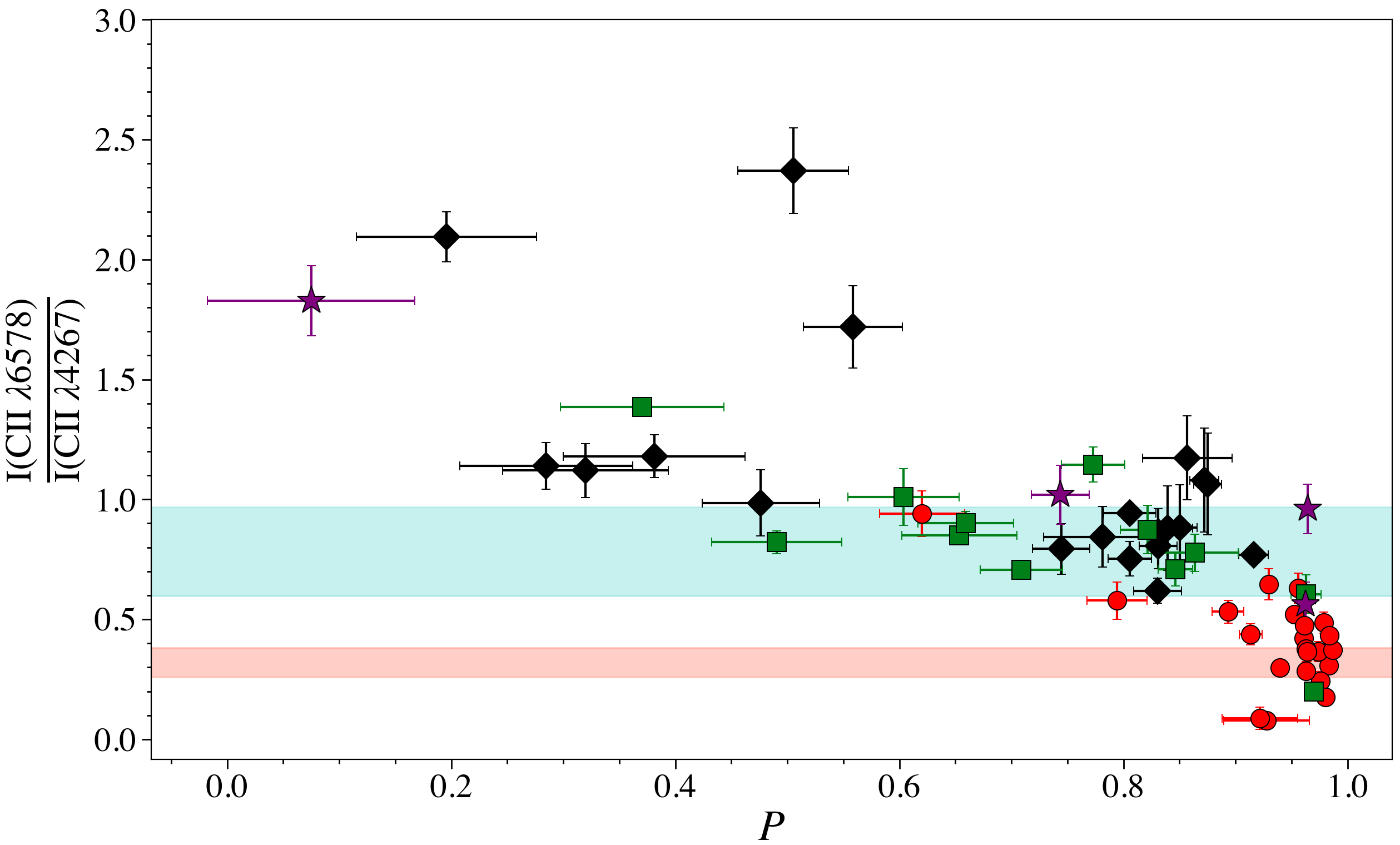}
\caption{I(\cii $\lambda$6578)/I(\cii $\lambda$4267) ratio of all analyzed regions with respect to the degree of ionization $P$. Symbols and color bands are the same as in Fig.~\ref{fig:tipos}.}
\label{fig:all1}
\end{figure}

Fig.~\ref{fig:all1} shows the I(\cii $\lambda$6578)/I({\cii}$\lambda$4267) line intensity ratio distribution with respect to $P$. This line ratio clearly shows a trend with $P$, indicating a significant fluorescent contribution to the excitation of {\cii} $\lambda$6578 in at least several nebulae. Additionally, there is a notable inconsistency between the theore\-tical values predicted by the atomic parameters of \citet{pequignotetal91} and those of \citet{daveyetal00}, which are connected with the same trend found in Sect.~\ref{subsec:3918_7231} for \cii $\lambda$7231+36+37.

At this point, we can draw two immediate conclusions: i) the dominant excitation mechanism of the \cii $\lambda$6578 line is not recombination, but continuum fluorescence in most cases; ii) using the theoretical recombination predictions for an optically thick nebula for this line seem inadequate.

\begin{figure}
\centering
    \includegraphics[width=\hsize]{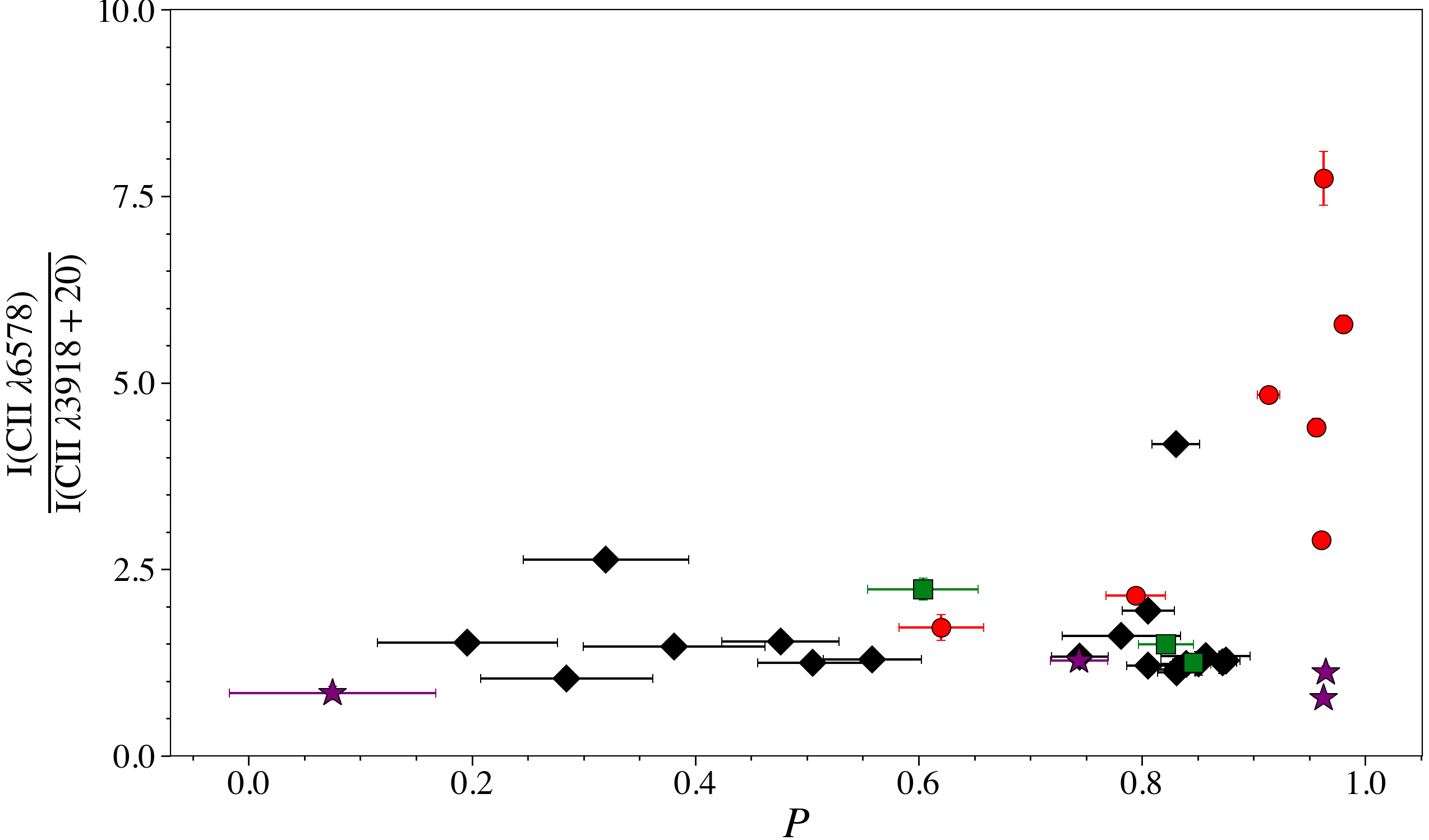}
    \includegraphics[width=\hsize]{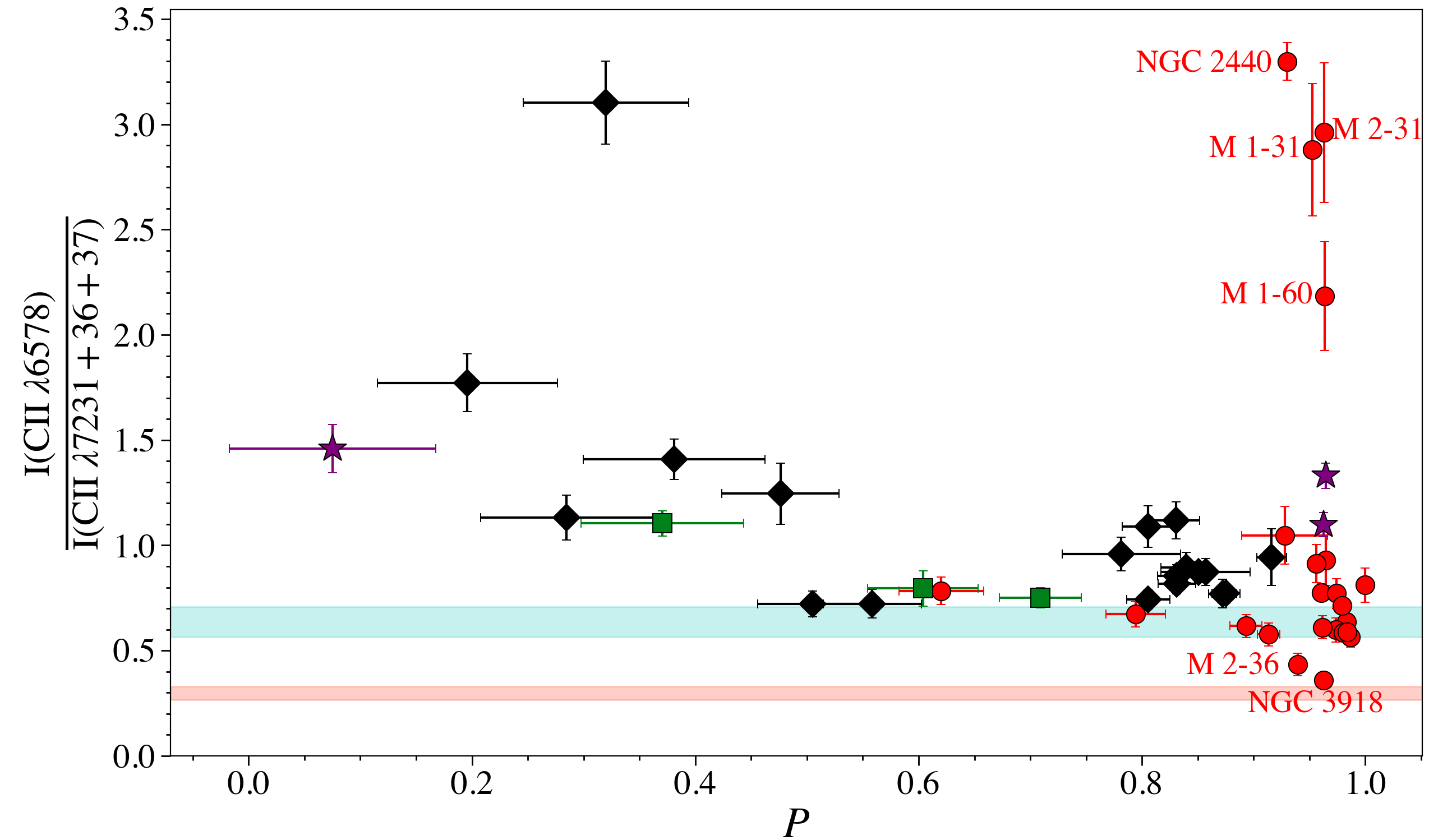}
\caption{I(\cii $\lambda$6578)/I(\cii $\lambda$3918+20) ratio (top panel) and  I(\cii $\lambda$6578)/I(\cii $\lambda$7231+36+37) ratio (bottom panel) for all analyzed regions with respect to the degree of ionization. Symbols and colorbands are the same as in Fig.~\ref{fig:tipos}.}
\label{fig:all2}
\end{figure}

As it has been shown that fluorescence is in most cases a key mechanism for accounting for the intensity of the \cii $\lambda$6578 line, we now aim at investigating the dominant cha\-nnel populating the $3p\ ^{2}P^{o}$ level. We first focus on cascade decays related with \cii $\lambda\lambda$3918, 3920 lines that po\-pu\-la\-te this level. The uppermost panel of Fig.~\ref{fig:all2} shows that for most objects the I(\cii $\lambda$6578)/I(\cii $\lambda$3918+20) ratio remains relatively constant, except for a small sample of highly-excited PNe where values larger than 10 are reached. This behaviour seems to indicate that for most ionized regions the \cii $\lambda$6578 emission primarily results from cascade decays via \cii $\lambda\lambda$3918, 3920 line emission. It is worth noting that two PNe (NGC\,2440 and M\,2-36) fall well above the bulk of the objects in the y-axis. Therefore, the large dispersion of the I(\cii $\lambda$6578)/I(\cii $\lambda$3918+20) ratio in PNe prevents us from being conclusive for these objects concerning the prevalence of cascade decays via the \cii $\lambda\lambda$3918, 3920 lines for populating the upper level of the \cii $\lambda6578$ line. 
It should be noted that the data set for \cii $\lambda\lambda$3918, 3920 lines is smaller due to their location in the bluest region of the optical spectrum and have rather faint intensities prior to the reddening correction, making their measurements challenging. Finally, we should also mention that the theoretical recombination predictions from \citet{daveyetal00} are out of the range of the observed I(\cii $\lambda$6578)/I(\cii $\lambda$3918+20) values. We could not compare with  predictions from \citet{pequignotetal91} as the authors did not make calculations of the  \cii $\lambda\lambda$3918, 3920 lines.

The other channel that can populate the $3p\ ^{2}P^{o}$ level is the cascade decay of the transition which gives rise to the \cii $\lambda\lambda$7231, 7236, 7237 lines, in which fluorescence competes with recombination for becoming the main excitation mechanism. Interestingly, it has been estimated by \citet{grandi76} that in the Orion Nebula the contribution  of either process is around 50\%. Similarly to what has been done before, we now exami\-ne the I(\cii $\lambda$6578)/I(\cii $\lambda$7231+36+37) line ratio in relation to the ionization degree. In the lower panel of Fig.~\ref{fig:all2}, we present the resulting line ratio versus $P$.
In this figure, we observe a possible dependence on $P$, contrary to what has been found with the I(\cii $\lambda$6578)/I(\cii $\lambda$3918+20) ratio, which seems to indicate that the primary fluorescence channel is the same as for the \cii $\lambda\lambda$3918, 3920 lines.

\begin{figure}
\centering
    \includegraphics[width=\hsize]{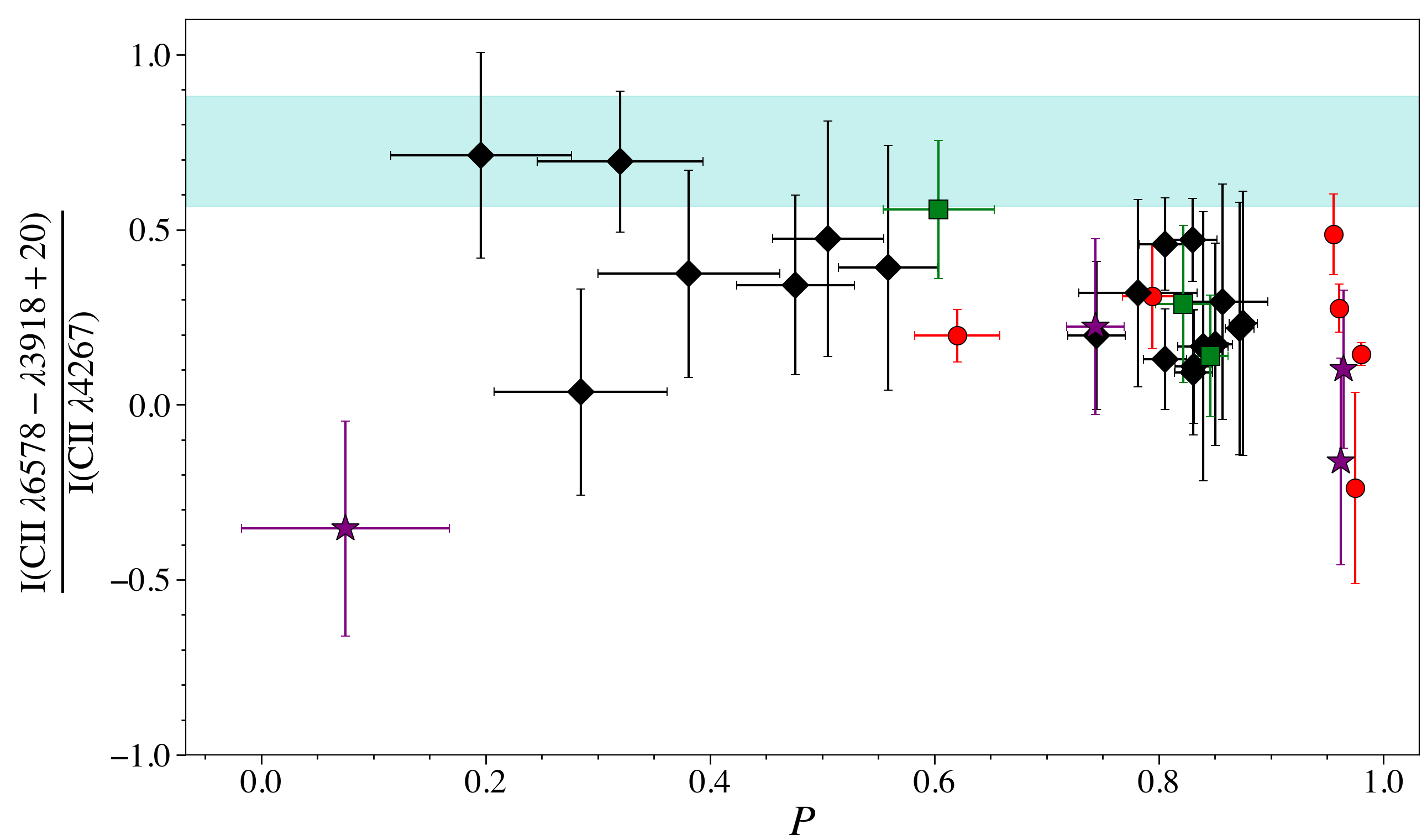}
\caption{I(\cii $\lambda$6578)-I(\cii $\lambda$3918+20)/I(\cii $\lambda$4267) ratio for all observed regions with respect to the degree of ionization $P$. Symbols and colorbands are the same as in Fig.~\ref{fig:tipos}. The blue band shows the theoretical ratios expected when adopting the atomic data of D00. Note that \citet{pequignotetal91} did not carry out calculations of \cii $\lambda$3918+20.}
\label{fig:resta1}
\end{figure}

A potential test to assess the impact of fluorescence via the \cii $\lambda \lambda$3918, 3920 channel for the nebulae of our sample is to subtract the intensity of these lines from the \cii $\lambda$6578 line and inspect the behaviour of the I(\cii $\lambda$6578)-I(\cii $\lambda$3918+20)/I(\cii $\lambda$4267) ratio by comparing with what was obtained in Fig.~\ref{fig:all1}. Fig.~\ref{fig:resta1} shows this ratio as a function of $P$  for all the regions in which the spectral lines involved were measured. We have also included the theoretical ratio expected for the same nebular conditions as in the previous cases (blue band), assuming the atomic data of D00. It is quite clear from this figure that for certain regions (mostly Galactic {\hii} regions due to the limited data available for the other type of regions), the fluorescence contribution of the channel that populates the upper level of the \cii $\lambda$$\lambda$3918, 3920 lines seems to have been eliminated, indicating that for these regions this channel is the dominant one. This is evident when one observes that the line intensity ratio with respect to \cii $\lambda$4267, a recombination line (see Sect.~\ref{subsec:recs_lines}), is nearly flat, independent of $P$.
Therefore, for {\hii} regions, the primary fluorescence channel populating the $3p\ ^{2}P^{o}$ level, from which the  \cii $\lambda$6578 line ori\-gi\-nates, appears to be that of the transitions that generate the \cii $\lambda\lambda$3918, 3920 lines. However, in the case of PNe, a more detailed analysis that could focus on each individual object is required, which lies however beyond the scope of current paper. Depending on the object, the fluorescence could originate from different channels and its contribution may not always be significant. In some cases, the predominant source of fluorescence may arise from the channel that causes the \cii $\lambda\lambda$7231, 7236, 7237 lines.

\section{Discussion}\label{sec:discuss}

It is well known that the C$^{2+}$/H$^{+}$ ionic abundances inferred from the \cii permitted lines are systematically higher than those derived from their collisionally excited counterparts: [\ciii] $\lambda 1907$, \ciii] $\lambda 1909$, when the ``direct method'' \citep{aller84,osterbrock88,Dinerstein:1990} is adopted \citep{torrespeimbertetal80,Walter:1992, Peimbert:1995, Esteban:1998}. To explain this inconsistency, some authors suggested that a fluorescent contribution to the excitation of the \cii\ permitted lines could explain at least part of the problem \citep{seaton68}. In the same vein, errors in the atomic parameters have also been invoked to explain this phenomenon \citep{Rodriguez:2010}.


Our results from Sect.~\ref{subsec:recs_lines} provide the observational e\-vi\-den\-ce to rule out that the cause of the abundance dis\-cre\-pan\-cy for the C$^{2+}$ ion, totally or partially, is due to fluorescence and/or errors in the atomic parameters of the permitted \cii lines $\lambda\lambda$4267, 5342, 6151, 6462 and 9903. Fig.~\ref{fig:tipos} and Table~\ref{tab:RLS_OBS_values} show excellent agreement between the observations and the predictions of the atomic models of \citet{pequignotetal91} and \citet{daveyetal00} for the aforementioned lines under the hypothesis of pure recombination. Furthermore, despite coming from different levels of the $^{2}F^{o} - ^{2}G$ and $^{2}D - ^{2}F^{o}$ transitions are insensitive to the ionization state of the gas, contrary to what should be observed if part of the emission of these lines comes from the fluorescent excitation of C$^+$. In fact, when high quality data is used to determine the C$^{2+}$/H$^{+}$ abundance with the aforementioned lines, a good consistency is always found \citep{garciarojasetal13}. These findings show that the C$^{2+}$ abundance discrepancy has its origin in a physical phenomenon.

Another observational evidence that reinforces that the origin of the C$^{2+}$ abundance discrepancy due to internal physical complexities of the ionized nebulae is that the C/O abundances, derived with both optical RLs and UV-CELs, vary from region to region and are consistent with each other \citep{torrespeimbertetal80, estebanetal05, estebanetal14, wangliu07, delgadoingladarodriguez14, Berg:2016, Izotov:2023}. The fact that C/O abundances from \hii\ regions and planetary nebulae, derived both with optical RLs and UV-CELs are consistent, could suggest that the abundance discrepancies of C$^{2+}$ and O$^{2+}$ are originated by the presence of temperature inhomogeneities and/or by the presence of metal-rich gas clumps.

An inhomogeneous temperature structure in the volume where C$^{2+}$ and O$^{2+}$ coexist can induce a systematic underestimation in the ionic abundances obtained with CELs, while those derived from RLs would not be affected \citep{peimbert67,peimbertcostero69,stasinska:1980,Kingdon:1995,garciarojasesteban07,Cameron:2023,mendezdelgado:2023}. Since [\oiii] and [\ciii] UV-CELs arise from atomic levels situated at similar energies, the temperature bias in both ions is rather similar and should therefore cancel out in the ionic abundance ratio $\text{C}^{2+}/\text{O}^{2+} \propto \text{C/O}$. On the other hand, if there are chemical inhomogeneities, as some authors have proposed to explain the AD problem, especially in the case of PNe \citep{torrespeimbertetal90,liuetal00,tsamisetal04,yuanetal11,gomezllanosmorisset20,garciarojasetal22}, most of the RL emissions will arise from metal-rich volumes as they would have a lower temperature. If these chemical inhomogeneities preserve the proportion of C and O (as in the case of H-deficient clumps), the C/O values obtained from RLs and CELs will be consistent, as is observed.

The recent observational evidence presented by \citet{mendezdelgado:2023} favors the presence of temperature inhomogeneities as the cause of the AD problem in \hii\ regions, as was originally proposed by \citet{peimbert67}. However, these authors show (see the Fig.~10 from the ArXiv version) that this mechanism does not act in the same way in PNe. In these objects, the abundance discrepancy might be caused mainly by the presence of chemical inhomogeneities \citep{torrespeimbertetal90,liuetal00,yuanetal11,garciarojasetal22}. However, as shown by \citet{richeretal22}, the presence of chemical inhomogeneities does not imply that temperature inhomogeneities are absent. The former phenomenon could in fact cause the latter \citep{Zhang:2007}. The new generation of photoionization models ought to consider these complexities, as is currently being proposed by \citet{Jin:2023} and \citet{Marconi:24}.

The analysis carried out in Sect.~\ref{subsec:3918_7231} emphasizes the importance of testing all theoretical models with dedicated observations. Although the excitation mechanisms of the \cii $\lambda \lambda$3918, 3920, 7231, 7236, 7237 lines had been previously analyzed in individual nebular regions \citep{grandi75a, escalanteetal12}, our results, which are based on a general sample of ionized nebulae, reveal that the fraction of excitation attributed to recombination in the case of the \cii $\lambda \lambda$7231, 7236, 7237 multiplet could be overestimated when one solely uses the atomic recombination parameters from \citet{pequignotetal91} and \citet{daveyetal00} to predict the recombination contribution to the observed line fluxes under the assumption of an optically thick nebula. This can furthermore generate important errors in interpretation of photoionization models not only for the aforementioned lines but for all those lines that arise from the \cii $3d\ ^2D$ level as well as those that arise from interconnection levels such as \cii $\lambda$6578. 

The case of the \cii $\lambda$6578 line is particularly interesting due to its proximity to the H$\alpha$ line and has been detected in a large number of objects. If this line was a pure recombination line, it could turn out being a useful tool in any study related to the AD problem. In fact, \citet{richeretal17} presented high spectral resolution spectroscopic observations of the \cii $\lambda$6578 line over 83 lines of sight across a sample of 76 PNe. These authors calculated a set of photoionization models with central stars (assumed to be a blackbody) that span the temperature range 40,000 K–150,000\,K in order to test the contribution of fluorescence to the emissivity of the \cii $\lambda$6578 line. They concluded that fluorescence may contribute to a significant fraction of the total emissivity of this line, especially in  models that assume a low-temperature for the central star (see their fig.~7, top row), but fluorescence never dominates the integrated surface brightness of the \cii  line. These kind of models depend on several assumptions on the effective temperatures of the ionizing stars as well as the accuracy of the effective recombination coefficients. As we show in Sect.~\ref{subsec:3918_7231} and Sect.~\ref{subsec:6578}, this is not straightforward as the fluorescence dominates the \cii $\lambda$6578 excitation thorough cascade decays that give rise to the \cii $\lambda$3918+20 lines. This unfortunate situation, however, complicates the use of the \cii $\lambda$6578 line when e\-va\-lua\-ting the recombination processes which regulate the C$^{2+}$ ion.\\

\section{Conclusions}
\label{sec:conclu} 

In this work, we have analyzed the excitation mechanisms of the permitted \cii lines $\lambda \lambda$3918, 3920, 4267, 5342, 6151, 6462, 7231, 7236, 7237 and 9903, widely detected on deep spectra in the spectral range around 3100-10400\AA. We use the DESIRED database \citep{mendezdelgado:2023b}, which contains many of the deepest spectra of photoionized regions of the literature (See Table.~\ref{tab:data} for references of the spectra), including \hii\ regions, PNe, and photoionized Herbig-Haro objects with a wide range of different physical and ionization conditions. Our main results on the excitation mechanisms of these lines can be summarized in Table~\ref{tab:summary}.

\setcounter{table}{4}
\begin{table*}
\centering
\scriptsize
\caption{\small Electronic configurations and dominant mechanism of the \cii transitions analysed in this work.}
\label{tab:summary}
\begin{tabular}{cccccccc}
    \hline
    Lab. wavelength [Å] & Spc & Configuration & Term & Dominant Mechanism \\   \hline
    4267.001 & {\cii} & 2s$^2$.3d-2s$^2$.4f & $^2$D-$^2$F$^{\mathrm{o}}$ & Recombination \\ 
    5342.500 & {\cii} & 2s$^2$.4f-2s$^2$.7g & $^2$F$^{\mathrm{o}}$-$^2$G & Recombination \\
    6151.530 & {\cii} & 2s$^2$.4d-2s$^2$.6f & $^2$D-$^2$F$^{\mathrm{o}}$ & Recombination \\
    6461.950 & {\cii} & 2s$^2$.4f-2s$^2$.6g & $^2$F$^{\mathrm{o}}$-$^2$G & Recombination \\
    9903.890 & {\cii} & 2s$^2$.4f-2s$^2$.5g & $^2$F$^{\mathrm{o}}$-$^2$G & Recombination \\
    3918.978 & {\cii} & 2s$^2$.3p-2s$^2$.4s & $^2$P$^{\mathrm{o}}$-$^2$S & Fluorescence  \\
    3920.693 & {\cii} & 2s$^2$.3p-2s$^2$.4s & $^2$P$^{\mathrm{o}}$-$^2$S & Fluorescence  \\
    6578.048 & {\cii} & 2s$^2$.3s-2s$^2$.3p & $^2$S-$^2$P$^{\mathrm{o}}$ & Fluorescence  \\
    7231.340 & {\cii} & 2s$^2$.3p-2s$^2$.3d & $^2$P$^{\mathrm{o}}$-$^2$D  & Fluorescence/Recombination \\
    7236.420 & {\cii} & 2s$^2$.3p-2s$^2$.3d & $^2$P$^{\mathrm{o}}$-$^2$D  & Fluorescence/Recombination \\
    7237.170 & {\cii} & 2s$^2$.3p-2s$^2$.3d & $^2$P$^{\mathrm{o}}$-$^2$D  & Fluorescence/Recombination \\
    \hline
    \end{tabular}
\end{table*}

Our methodology is robust and straightforward since it is observationally-based and only uses the initial hypothesis that the emissivity of fluorescent permitted \cii lines will be proportional to the C$^{+}$ abundance, whereas those arising from recombinations will depend on the  C$^{2+}$ abundance. The weak dependence of the emissivity of the \cii permitted lines on the physical conditions ($T_{\rm e}$, $n_{\rm e}$) allow us to directly address their excitation mechanisms and the accuracy of the atomic models from \citet{pequignotetal91} and  \citet{daveyetal00}, the most widely used to infer C$^{2+}$/H$^{+}$ abundances and photoionization modeling.

We show that \cii $\lambda \lambda$4267, 5342, 6151, 6462 and 9903 are produced by pure recombination. We also show that the recombination theory from \citet{pequignotetal91} and \citet{daveyetal00} for these lines is essentially correct. This implies that the long-standing abundance discrepancy problem between the C$^{2+}$/H$^{+}$ abundances derived with the aforementioned lines and their UV collisionally excited counterparts ([\ciii] $\lambda 1907$ and \ciii] $\lambda1909$) is produced by a physical phenomenon other than fluorescence or errors in atomic recombination coefficients.

On the other hand, we find that \cii $\lambda\lambda$3918, 3920 lines are excited mainly by fluorescence, while \cii $\lambda \lambda$7231, 7236, 7237 lines have important contributions from both fluorescence and recombinations. We leave open the possibility of having some overestimation of the recombination emissivity of the \cii $\lambda \lambda$7231, 7236, 7237 lines by both the atomic parameters of \citet{pequignotetal91} and \citet{daveyetal00}, which can induce systematic errors in the photoionization modelling of these lines. Photoionization models with detailed information about the ionizing stars of a broad sample of nebulae will shed light on this possibility. Finally, we demonstrate that the main excitation mechanism of \cii $\lambda$6578 is continuum fluorescence in most cases.

\begin{acknowledgements}
      JEM-D and KK gratefully acknowledges funding from the Deutsche Forschungsgemeinschaft (DFG, German Research Foundation) in the form of an Emmy Noether Research Group (grant number KR4598/2-1, PI Kreckel) and the European Research Council's starting grant ERC StG-101077573 ("ISM-METALS"). JG-R acknowledges financial support from the Canarian Agency for Research, Innovation and Information Society (ACIISI), of the Canary Islands Government, and the European Regional Development Fund (ERDF), under grant with reference ProID2021010074, and from grant P/308614 financed by funds transferred from the Spanish Ministry of Science, Innovation and Universities, charged to the General State Budgets and with funds transferred from the General Budgets of the Autonomous Community of the Canary Islands by the MCIU. JG-R also acknowledges funds from the Spanish Ministry of Science and Innovation (MICINN) through the Spanish State Research Agency, under Severo Ochoa Centres of Excellence Program 2020-2023 (CEX2019-000920-S). We acknowledge support from the Agencia Estatal de Investigación del Ministerio de Ciencia e Innovación (AEI- MCINN) under grant Espectroscopía de campo integral de regiones H II locales. Modelos para el estudio de regiones H II extragalácticas with reference 10.13039/501100011033.

\end{acknowledgements}

%
%

\begin{appendix} 

\onecolumn
\section{Spectra in the DESIRED database with detection of at least two permitted \cii lines.}

\centering
\setcounter{table}{0}
\begin{scriptsize}
\setlength{\tabcolsep}{4pt}
\begin{longtable}{ccccccccccc}
        \caption{\label{tab:data} Extinction-corrected $I$($\lambda$)/$I$(\hb) (in units of $I$(\hb) = 100.0) of {\cii} lines detected in the spectra of the DESIRED database.}
        \tabularnewline
        \hline
        \hline
        \multicolumn{11}{ c }{\textbf{Galactic Planetary Nebulae}} \\        
        \hline        
        \textbf{Name}  & \textbf{$\lambda$3918+20} & \textbf{$\lambda$4267} & \textbf{$\lambda$5342} & \textbf{$\lambda$6151} & \textbf{$\lambda$6462} & \textbf{$\lambda$6578} & \textbf{$\lambda$7231+36+37} & \textbf{$\lambda$9903} & $P$ & Ref. \\
        \hline
        \endfirsthead
        \caption{continued.}\tabularnewline
        \hline
        \endhead
        \endfoot
        \endlastfoot
        Abell\,46 & {\it -} & 6.6$\pm$0.3 & {\it -} &{\it -} & 0.47$\pm$0.12 & 0.58$\pm$0.12 & {\it -} & {\it -} & 0.92$\pm$0.03 & 1 \\

        Cn\,1-5 & 0.19$\pm$0.08 & 1.4$\pm$0.1 & 0.099$\pm$0.015 & 0.058$\pm$0.012 & {\it -} & 0.61$\pm$0.04 & 1.06$\pm$0.08 & {\it -} & 0.913$\pm$0.010 & 2 \\

        H\,1-40 & {\it -} & {\it -} & {\it -} & {\it -} & {\it -} & 0.13$\pm$0.04 & 0.14$\pm$0.03 & 0.049$\pm$0.009 & 0.913$\pm$0.010 &  3 \\

        H\,1-50 & {\it -} & 0.35$\pm$0.05 & {\it -} & {\it -} & {\it -} & 0.084$\pm$0.018 & {\it -} & 0.084$\pm$0.009  & 0.964$\pm$0.008 & 3 \\

        Hb\,4 & {\it -} & 0.76$\pm$0.09 & 0.027$\pm$0.011 & 0.025$\pm$0.007 & 0.08$\pm$0.01 & 0.23$\pm$0.03 & 0.37$\pm$0.05 & {\it -} & 0.983$\pm$0.003 & 2 \\

        Hen\,2-73 & {\it -} & 0.59$\pm$0.07 & {\it -} & {\it -} & {\it -} & {\it -} & 0.26$\pm$0.02 & 0.144$\pm$0.009 & 0.966$\pm$0.004 & 3 \\

        Hen\,2-86 & {\it -} & 0.73$\pm$0.05 & 0.042$\pm$0.004 & 0.028$\pm$0.003 & 0.066$\pm$0.006 & 0.32$\pm$0.03 & 0.54$\pm$0.06 & {\it -}  & 0.984$\pm$0.003 & 2 \\

        Hen\,2-96 & {\it -} & 0.36$\pm$0.06 & {\it -} & {\it -} & {\it -} & {\it -} & 0.25$\pm$0.03 & 0.082$\pm$0.006 & 0.958$\pm$0.007  & 3 \\

        IC\,418 & 0.312$\pm$0.009 & 0.571$\pm$0.017 & 0.028$\pm$0.003 & 0.025$\pm$0.003 & 0.058$\pm$0.006 & 0.537$\pm$0.016 & 0.69$\pm$0.02 & {\it -} & 0.62$\pm$0.04  & 4 \\

        IC\,2501 & 0.143$\pm$0.016 & 1.00$\pm$0.03 & 0.050$\pm$0.006 & 0.041$\pm$0.005 & 0.110$\pm$0.003 &  0.630$\pm$0.019 & 0.69$\pm$0.02 &  {\it -} & 0.956$\pm$0.004  & 5 \\

        IC\,4191 & 0.019$\pm$0.003 & 0.630$\pm$0.019 & 0.030$\pm$0.003 & {\it -} & {\it -} & 0.011$\pm$0.003 & 0.188$\pm$0.011 & {\it -} & 0.980$\pm$0.002  &  5 \\

        IC\,4776 & 0.039$\pm$0.009 & 0.164$\pm$0.003 & 0.010$\pm$0.002 & 0.009$\pm$0.001 & 0.018$\pm$0.002 & {\it -} & 0.094$\pm$0.006 & {\it -} & 0.975$\pm$0.006 & 6 \\

        M\,1-25 & {\it -} & 0.59$\pm$0.06 & 0.036$\pm$0.005 & 0.024$\pm$0.003 & 0.062$\pm$0.006 & 0.31$\pm$0.03 & 0.51$\pm$0.05 &  {\it -} & 0.893$\pm$0.014 &  2 \\

        M\,1-30 & 0.26$\pm$0.04 & 0.98$\pm$0.06 & 0.055$\pm$0.006 & 0.035$\pm$0.004 & 0.106$\pm$0.008 & 0.57$\pm$0.05 & 0.84$\pm$0.08 &  {\it -} & 0.79$\pm$0.03 & 2 \\

        M\,1-31 & {\it -} & 0.60$\pm$0.13 & {\it -} & {\it -} & 0.07$\pm$0.02 & 0.31$\pm$0.04 & 0.109$\pm$0.012 & 0.124$\pm$0.009 & 0.952$\pm$0.007 &  3 \\

        M\,1-32 & {\it -} & 2.03$\pm$0.16 & 0.16$\pm$0.03 & 0.070$\pm$0.018 & {\it -} & {\it -} & 2.2$\pm$0.3 & {\it -} & 0.84$\pm$0.03 &  2 \\

        M\,1-33 & {\it -} & 1.05$\pm$0.09 & {\it -} & {\it -} & 0.10$\pm$0.03 & {\it -} & 0.52$\pm$0.04 &  0.236$\pm$0.014 & 0.957$\pm$0.005 & 3 \\

        M\,1-60 & {\it -} & 0.88$\pm$0.10 & {\it -} & {\it -} & 0.09$\pm$0.03 & 0.32$\pm$0.04 & 0.147$\pm$0.016 & 0.197$\pm$0.014 & 0.964$\pm$0.005 & 3 \\

        M\,1-61 & {\it -} & 0.45$\pm$0.05 & 0.028$\pm$0.006 & 0.021$\pm$0.004 & 0.047$\pm$0.006 & 0.211$\pm$0.019 & 0.31$\pm$0.03 & {\it -} & 0.979$\pm$0.003 & 2 \\

        M\,2-31 & {\it -} & 0.53$\pm$0.08 & {\it -} & {\it -} & {\it -} & 0.15$\pm$0.03 & 0.051$\pm$0.009 & 0.120$\pm$0.008 & 0.963$\pm$0.005 &  3 \\

        M\,2-36 & 0.051$\pm$0.016 & 2.34$\pm$0.12 & 0.116$\pm$0.019 & 0.072$\pm$0.009 & 0.250$\pm$0.015 & 0.70$\pm$0.03 & 1.62$\pm$0.05 &  {\it -} & 0.940$\pm$0.007 & 7 \\

        M\,3-15 & {\it -} & 0.55$\pm$0.16 & 0.047$\pm$0.013 & 0.038$\pm$0.008 & 0.077$\pm$0.012 & 0.20$\pm$0.02 & 0.36$\pm$0.05 & {\it -} & 0.987$\pm$0.003 & 2 \\

        NGC\,2440 & 0.012$\pm$0.001 & 0.480$\pm$0.014 & 0.021$\pm$0.002 & {\it -} & {\it -} & 0.310$\pm$0.009 & 0.094$\pm$0.010 & {\it -} & 0.930$\pm$0.007 & 5 \\

        NGC\,3918 & 0.023$\pm$0.003 & 0.47$\pm$0.02 & 0.025$\pm$0.003 & 0.021$\pm$0.003 & 0.049$\pm$0.004 & 0.178$\pm$0.012 & 0.49$\pm$0.04 &  {\it -} & 0.963$\pm$0.004 & 8 \\

        NGC\,5189 & {\it -} & 0.30$\pm$0.04 & {\it -} & 0.022$\pm$0.007 & {\it -} & {\it -} & 0.067$\pm$0.015 & {\it -} & 0.866$\pm$0.018 & 2 \\
         
        NGC\,5315 & 0.10$\pm$0.02 & 0.72$\pm$0.06 & 0.035$\pm$0.004 & 0.027$\pm$0.003 & 0.066$\pm$0.005 & 0.301$\pm$0.018 & 0.39$\pm$0.04 & {\it -} & 0.961$\pm$0.006 & 9 \\

        NGC\,6369 & {\it -} & 0.79$\pm$0.07 & {\it -} & 0.045$\pm$0.011 & 0.092$\pm$0.012 & 0.29$\pm$0.02 & 0.49$\pm$0.04 & {\it -} & 0.974$\pm$0.003 & 2 \\

        NGC\,6778 & {\it -} & 1.86$\pm$0.07 & {\it -} & {\it -} & 0.17$\pm$0.03 & {\it -} & {\it -} & {\it -} &  0.90$\pm$0.04 & 10 \\

        Ou5 & {\it -} & 5.7$\pm$0.7 & {\it -} & {\it -} & 0.38$\pm$0.13 & 0.44$\pm$0.13 & 0.42$\pm$0.09 & {\it -} & 0.93$\pm$0.04  & 1 \\

        PC\,14 & {\it -} & 0.88$\pm$0.06 & 0.037$\pm$0.013 & 0.032$\pm$0.011 & 0.084$\pm$0.012 & 0.32$\pm$0.03 & 0.42$\pm$0.04 & {\it -} & 0.974$\pm$0.003 & 2 \\

        Pe\,1-1 & {\it -} & 1.12$\pm$0.09 & 0.058$\pm$0.010 & 0.050$\pm$0.007 & 0.105$\pm$0.011 & 0.53$\pm$0.05 & 0.87$\pm$0.09 & {\it -} & 0.961$\pm$0.005 & 2 \\

        \hline
        \hline
        \multicolumn{11}{ c }{\textbf{Galactic \hii\   regions}}\\        
        \hline
        \textbf{Name}  & \textbf{$\lambda$3918+20} & \textbf{$\lambda$4267} & \textbf{$\lambda$5342} & \textbf{$\lambda$6151} & \textbf{$\lambda$6462} & \textbf{$\lambda$6578} & \textbf{$\lambda$7231+36+37} & \textbf{$\lambda$9903} & $P$ & Ref. \\
        \hline

        M8 & 0.179$\pm$0.018 &  0.222$\pm$0.009 & 0.011$\pm$0.004 & {\it -} & 0.025$\pm$0.004 & 0.262$\pm$0.008 &  0.186$\pm$0.008 & 0.048$\pm$0.003 & 0.38$\pm$0.08 & 11 \\

        M16 & 0.30$\pm$0.04 & 0.272$\pm$0.019 & {\it -} & {\it -} & 0.032$\pm$0.012 & 0.310$\pm$0.019 & 0.274$\pm$0.017 & 0.037$\pm$0.005 & 0.28$\pm$0.08 & 12 \\

        M17 & 0.09$\pm$0.02 & 0.58$\pm$0.04 & {\it -} & {\it -} & 0.050$\pm$0.007 &  0.359$\pm$0.018 & 0.32$\pm$0.02 & {\it -} & 0.83$\pm$0.02 & 11 \\

        M20 & 0.24$\pm$0.04 & 0.17$\pm$0.02 & {\it -} & {\it -} & {\it -} & 0.36$\pm$0.02 & 0.201$\pm$0.016 & {\it -} & 0.20$\pm$0.08 & 12 \\ 

        M42 & 0.218$\pm$0.016 & 0.248$\pm$0.010 & 0.013$\pm$0.004 & 0.008$\pm$0.003 & 0.025$\pm$0.004 & 0.291$\pm$0.017 & 0.33$\pm$0.03 & 0.059$\pm$0.009 & 0.86$\pm$0.04 & 13 \\

        M42 & 0.15$\pm$0.02 & 0.25$\pm$0.02 & {\it -} & {\it -} & {\it -} & 0.20$\pm$0.02 & {\it -} & 0.048$\pm$0.019 & 0.74$\pm$0.03 & 14 \\

        M42-1  & 0.149$\pm$0.009 & 0.239$\pm$0.012 & {\it -} & {\it -} & 0.026$\pm$0.005 & 0.180$\pm$0.009 & 0.242$\pm$0.019 & 0.056$\pm$0.006 & 0.806$\pm$0.019 & 15 \\

        M42-1 & 0.256$\pm$0.009 & 0.135$\pm$0.005 & {\it -} & {\it -} & {\it -} & 0.320$\pm$0.010 & 0.444$\pm$0.016 & 0.045$\pm$0.005 & 0.51$\pm$0.05  & 16 \\

        M42-2  & 0.165$\pm$0.005 & 0.237$\pm$0.008 & 0.016$\pm$0.002 & {\it -} & 0.022$\pm$0.003 & 0.191$\pm$0.008 & 0.234$\pm$0.017 & 0.061$\pm$0.005 & 0.831$\pm$0.017 & 15 \\

        M42-2  & 0.336$\pm$0.018 & 0.253$\pm$0.010 & {\it -} & {\it -} & {\it -} & 0.435$\pm$0.017 & 0.60$\pm$0.03 & 0.057$\pm$0.008 & 0.56$\pm$0.04 & 16 \\

        M42-2  & 0.214$\pm$0.007 & 0.248$\pm$0.005 & 0.011$\pm$0.004 & 0.014$\pm$0.003 & 0.023$\pm$0.002 & 0.268$\pm$0.011 & 0.348$\pm$0.016 & 0.065$\pm$0.006 & 0.872$\pm$0.013 & 17 \\

        M42-3  & 0.165$\pm$0.008 & 0.232$\pm$0.007 & 0.018$\pm$0.003 & 0.012$\pm$0.003 & 0.023$\pm$0.003 & 0.205$\pm$0.012 & 0.24$\pm$0.03 & 0.053$\pm$0.007 & 0.851$\pm$0.015 & 15 \\

        M42-3  & 0.207$\pm$0.008 & 0.249$\pm$0.005 & 0.016$\pm$0.004 & {\it -} & 0.024$\pm$0.003 &  0.265$\pm$0.011 & 0.34$\pm$0.03 & 0.065$\pm$0.006 & 0.875$\pm$0.012 &  17 \\

        M42-4  & 0.175$\pm$0.008 & 0.227$\pm$0.010 & {\it -} & {\it -} & 0.025$\pm$0.004 & 0.196$\pm$0.010 & 0.229$\pm$0.019 & 0.060$\pm$0.006 &  0.831$\pm$0.017 & 15 \\

        M42-bar & 0.126$\pm$0.012 & 0.196$\pm$0.010 & {\it -} & {\it -} & 0.017$\pm$0.006 & 0.193$\pm$0.014 & 0.155$\pm$0.009 & 0.045$\pm$0.012 & 0.48$\pm$0.05 &  18 \\

        M42-P1  & 0.18$\pm$0.02 & 0.252$\pm$0.010 & {\it -} & {\it -} & 0.027$\pm$0.007 & 0.222$\pm$0.018 & 0.25$\pm$0.03 & 0.061$\pm$0.011 & 0.84$\pm$0.02 &  18 \\

        NGC\,2579 & 0.076$\pm$0.017 & 0.157$\pm$0.019 & {\it -} & {\it -} & {\it -} & 0.148$\pm$0.007 & 0.136$\pm$0.008 & 0.032$\pm$0.002 & 0.81$\pm$0.02 & 19 \\

        NGC\,3576 & 0.155$\pm$0.017 & 0.295$\pm$0.012 & {\it -} & 0.012$\pm$0.004 & 0.032$\pm$0.005 & 0.249$\pm$0.015 & 0.260$\pm$0.019 & 0.080$\pm$0.009 & 0.78$\pm$0.05 & 20 \\

        NGC\,3603 & {\it -} & 0.33$\pm$0.06 & {\it -} & {\it -} & {\it -} & 0.25$\pm$0.02 & 0.265$\pm$0.017 & 0.111$\pm$0.010 & 0.916$\pm$0.013 & 12 \\

        Sh\,2-152 & 0.39$\pm$0.13 &  0.10$\pm$0.04 & {\it -} & {\it -} & {\it -} & {\it -} & 0.22$\pm$0.06 & {\it -} & 0.18$\pm$0.08 & 21 \\

        Sh\,2-311 & 0.046$\pm$0.012 & 0.108$\pm$0.013 & {\it -} & {\it -} & {\it -} & 0.121$\pm$0.015 & 0.039$\pm$0.007 & {\it -} & 0.32$\pm$0.07 & 22 \\
        
        \hline
        \hline
        \multicolumn{11}{ c }{\textbf{Extragalactic \hii\   regions}}\\        
        \hline
        \textbf{Name}  & \textbf{$\lambda$3918+20} & \textbf{$\lambda$4267} & \textbf{$\lambda$5342} & \textbf{$\lambda$6151} & \textbf{$\lambda$6462} & \textbf{$\lambda$6578} & \textbf{$\lambda$7231+36+37} & \textbf{$\lambda$9903} & $P$ & Ref. \\
        \hline

        30 Doradus & 0.053$\pm$0.010 & 0.093$\pm$0.009 & {\it -} & 0.023$\pm$0.003 & {\it -} & 0.066$\pm$0.007 & {\it -} & {\it -} & 0.846$\pm$0.015 &  23 \\

        H1013g & {\it -} & 0.28$\pm$0.08 & {\it -} & {\it -} & {\it -} & 0.23$\pm$0.04 & {\it -} & {\it -} & 0.49$\pm$0.06 & 24 \\
        
        IC\,2111a & 0.047$\pm$0.010 & 0.104$\pm$0.015 & {\it -} & {\it -} & {\it -} & 0.105$\pm$0.015 & 0.132$\pm$0.018 & {\it -} & 0.60$\pm$0.05 &  25 \\

        N81b & {\it -} & 0.036$\pm$0.002 & {\it -} & {\it -} & {\it -} & 0.028$\pm$0.002 & {\it -} & {\it -} & 0.86$\pm$0.04 & 25 \\

        N88Ab & {\it -} & 0.038$\pm$0.001 & {\it -} & {\it -} & {\it -} & 0.023$\pm$0.001 & {\it -} & 0.009$\pm$0.001 & 0.963$\pm$0.013 & 25 \\
        
        NGC\,595 & {\it -} & 0.13$\pm$0.03 & {\it -} & {\it -} & {\it -} & 0.180$\pm$0.011 & 0.163$\pm$0.018 & {\it -} & 0.37$\pm$0.07 &  24 \\

        NGC\,604 & {\it -} & 0.17$\pm$0.04 & {\it -} & {\it -} & {\it -} & 0.120$\pm$0.010 & 0.16$\pm$0.02 & {\it -} & 0.70$\pm$0.04 & 24 \\

        NGC\,1714a & 0.065$\pm$0.018 & 0.111$\pm$0.014 & {\it -} & {\it -} & {\it -} & 0.097$\pm$0.015 & {\it -} & {\it -} & 0.82$\pm$0.02 & 25 \\

        NGC\,2363 & {\it -} & 0.035$\pm$0.007 & {\it -} & {\it -} & {\it -} & 0.007$\pm$0.002 & {\it -} & {\it -} &  0.970$\pm$0.004 & 24 \\

        NGC\,5471 & {\it -} & 0.054$\pm$0.016 & {\it -} & {\it -} & {\it -} & {\it -} &  0.09$\pm$0.03 & {\it -} & 0.910$\pm$0.013 & 26 \\

        UV-1 & {\it -} & 0.06$\pm$0.03 & {\it -} & {\it -} & {\it -} & 0.071$\pm$0.019 & {\it -} & {\it -} & 0.77$\pm$0.03 & 27 \\

        VS\,38d & {\it -} & 0.20$\pm$0.06 & {\it -} & {\it -} & {\it -} & 0.17$\pm$0.02 & {\it -} & {\it -} & 0.65$\pm$0.05 & 24 \\

        VS\,44d & {\it -} & 0.10$\pm$0.04 & {\it -} & {\it -} & {\it -} & 0.09$\pm$0.02 & {\it -} & {\it -} & 0.66$\pm$0.04 & 24 \\

        \hline
        \hline
        \multicolumn{11}{ c }{\textbf{Photoionized Herbig-Haro objects}}\\        
        \hline
        \textbf{Name}  & \textbf{$\lambda$3918+20} & \textbf{$\lambda$4267} & \textbf{$\lambda$5342} & \textbf{$\lambda$6151} & \textbf{$\lambda$6462} & \textbf{$\lambda$6578} & \textbf{$\lambda$7231+36+37} & \textbf{$\lambda$9903} & $P$ & Ref. \\
        \hline
        HH202S & 0.17$\pm$0.03 & 0.21$\pm$0.02 & {\it -} & {\it -} & {\it -} & 0.22$\pm$0.03 & {\it -} & 0.045$\pm$0.018 & 0.74$\pm$0.03 &  14 \\
        
        HH204 & 0.216$\pm$0.008 & 0.099$\pm$0.005 & {\it -} & {\it -} & {\it -} & 0.181$\pm$0.007 & 0.124$\pm$0.006 & 0.015$\pm$0.003 & 0.08$\pm$0.09 & 16 \\

        HH529II & 0.263$\pm$0.017 & 0.306$\pm$0.014 & {\it -} & {\it -} & {\it -} & 0.294$\pm$0.015 & 0.22$\pm$ 0.02 & 0.078$\pm$0.012 & 0.965$\pm$0.004 & 15 \\

        HH529III & 0.28$\pm$0.05 & 0.38$\pm$ 0.02 & {\it -} & {\it -} & {\it -} & 0.21$\pm$ 0.02 & 0.20$\pm$ 0.04 & 0.11$\pm$0.03 & 0.962$\pm$0.004 & 15 \\
        \hline    
\end{longtable}
\end{scriptsize}
\tablebib{(1) \citet{corradietal15}; (2) \citet{garciarojasetal12}; (3) \citet{garciarojasetal18}; (4) \citet{sharpeeetal03}; (5) \citet{sharpeeetal07}; (6) \citet{sowickaetal17}; (7) \citet{espiritupeimbert21}; (8) \citet{garciarojasetal15}; (9) \citet{madonnaetal17}; (10) \citet{jonesetal16}; (11) \citet{garciarojasetal07}; (12) \citet{garciarojasetal06}; (13) \cite{estebanetal04}; (14) \cite{mesadelgadoetal09a}; (15) \cite{mendezdelgadoetal21}; (16) \cite{mendez-delgado21b}; (17) \cite{mendez-delgado22b}; (18) \cite{delgadoingladaetal16}; (19) \citet{estebanetal13}; (20) \citet{garciarojasetal04}; (21) \citet{estebangarciarojas18}; (22) \citet{garciarojasetal05}; (23) \citet{peimbert03}; (24) \citet{estebanetal09}; (25) \citet{dominguezguzmanetal22}; (26) \citet{Esteban:2020}; (27) \citet{lopezsanchezetal07}
}
\\

\end{appendix} 

\end{document}